\documentclass[twocolumn]{aastex631}

\usepackage{times}
\usepackage{amsmath,amssymb,amstext}
\hypersetup{linkcolor=red,citecolor=blue,filecolor=cyan,urlcolor=blue}
\usepackage{tabularx}
\usepackage{longtable}
\usepackage{float}
\usepackage{natbib}
\usepackage{graphicx,color}
\usepackage{txfonts}

\usepackage{multirow}
\usepackage{array}
\usepackage{rotating}
\usepackage{sidecap}
\usepackage{hyperref}
\usepackage{epstopdf}
\usepackage{footnote}
\usepackage{tabularx}
\usepackage{booktabs}
\usepackage{longtable}
\usepackage{tabu}
\usepackage{longtable}

\newcommand{\kms}{km\,s$^{-1}$}

\begin{document}

\title{{\large {Breakout/Interchange Reconnection as a driver of Jets, Fast Coronal Mass Ejection, and Solar Energetic Particles}}}
\author{Pankaj Kumar\altaffiliation{1,2}}

\affiliation{Department of Physics, American University, Washington, DC 20016, USA}
\affiliation{Heliophysics Science Division, NASA Goddard Space Flight Center, Greenbelt, MD, 20771, USA}

\author{Judith T.\ Karpen}
\affiliation{Heliophysics Science Division, NASA Goddard Space Flight Center, Greenbelt, MD, 20771, USA}

\author{David Lario}
\affiliation{Heliophysics Science Division, NASA Goddard Space Flight Center, Greenbelt, MD, 20771, USA}

\author{Benjamin J. Lynch}
\affiliation{Space Sciences Laboratory, University of California, Berkeley, CA 94720, USA}

\author{Peter F. Wyper}
\affiliation{Department of Mathematical Sciences, Durham University, Durham DH1 3LE, UK}

\author{Spiro K. Antiochos}
\affiliation{Department of Climate and Space Sciences and Engineering, University of Michigan, Ann Arbor, MI 48109, USA}
\affiliation{Heliophysics Science Division, NASA Goddard Space Flight Center, Greenbelt, MD, 20771, USA}

\author{P. K.\ Manoharan}
\affiliation{The Catholic University of America, 20664, Washington, DC, USA}
\affiliation{Heliophysics Science Division, NASA Goddard Space Flight Center, Greenbelt, MD, 20771, USA}

\email{pankaj.kumar@nasa.gov}

%*****************************************************************************
\begin{abstract}
Understanding how energetic particles are accelerated and released from the low corona into the interplanetary medium during solar eruptions is crucial for space weather research. Here, we present multiwavelength observations of a solar eruption that are consistent with breakout reconnection playing an important role in driving a fast coronal mass ejection (CME) and the associated solar energetic particle (SEP) event. Extreme-ultraviolet and radio observations reveal evidence of breakout reconnection within a fan–spine topology. The filament eruption begins after the pre-eruption opening, accompanied by quasiperiodic jets and associated downflows near the null point, as well as recurrent faint Type III radio bursts during the ongoing slow breakout reconnection. Furthermore, the observations also reveal the formation of pre-eruption coronal rain via slow interchange reconnection near the null point, a process with important implications for coronal heating and the origin of the solar wind. A large-scale circular ribbon, along with simultaneous four hard X-ray footpoint sources and intense Type III radio bursts, was observed during the explosive breakout reconnection that enabled opening of field lines and allowed energetic particles to escape into interplanetary space. In situ measurements by PSP and Wind confirm the prompt injection of electron beams consistent with the timing of the explosive breakout reconnection. A fast shock associated with the erupting flux rope during interchange reconnection played a major role in producing the gradual SEP event. These observations highlight the key role of breakout reconnection in producing a fast CME/shock with the SEP release and acceleration process. 
 These results have broader implications for particle acceleration and release processes in multiscale null-point topologies, which produce a continuum of eruptions ranging from small-scale jets to large-scale CMEs. 
  \end{abstract}
%*****************************************************************************
%*****************************************************************************
%%%%%% Section 1 %%%%%%%%%%%%%%%%%%%%%%%%%%%%%%%%%%%%%%%%%%%%%%%%%%%%%%%%%%

\section{INTRODUCTION}\label{intro}
Coronal pseudostreamers are large-scale magnetic structures that observationally appear as cusp-shaped arcades separating coronal holes of the same magnetic polarity \citep{wang2007}. In the topological sense, pseudostreamers correspond to multipolar magnetic configurations containing one or more coronal null points with associated fan--spine structures and separatrix features, including separators and quasi-separatrix layers \citep{titov2011,masson2014}. The presence of magnetic null points in pseudostreamers makes them natural sites for magnetic reconnection and energy release. Their large spatial scale and magnetic connectivity to open flux render them particularly important for understanding how energy stored in closed solar magnetic systems is transferred into the heliosphere.

Null-point topologies host a wide range of eruptive phenomena, from small-scale coronal jets (see review by \citealt{raouafi2016}, \citealt{sterling2015}, \citealt{kumar2018}) to large-scale coronal mass ejections (CMEs), with or without filament eruptions \citep{kumar2021}. Jets in such configurations are commonly interpreted within the breakout framework, where reconnection at a breakout current sheet (BCS) formed near the coronal null facilitates energy release in fan--spine magnetic structures \citep{antiochos1999, pariat2009, karpen2017, wyper2017, wyper2018a}. Magnetohydrodynamic simulations \citep{wyper2017} show that slow flare reconnection at the flare current sheet (FCS) beneath the rising core field contributes to flux-rope formation along the filament channel, while gradual breakout reconnection at the null removes the overlying strapping field. During the explosive phase, rapid reconnection at the BCS (above the flux rope) can drive reconnection jets accompanied by circular or remote ribbons, whereas reconnection at the FCS (below the flux rope) produces a typical two-ribbon flare signature. 
Observational studies have reported clear signatures of breakout reconnection at null points, including plasmoid formation within the BCS and the development of multiscale, sometimes quasiperiodic, jets with or without associated filament eruptions \citep{kumar2018,kumar2019a,kumar2019b,kumar2022,kumar2024}. Breakout reconnection, a subset of interchange reconnection operating in multipolar magnetic configurations, is adopted here to describe the energy-release process, and this term is used consistently throughout the paper.

We demonstrated that eruptions from pseudostreamer magnetic configurations often begin with subtle pre-eruption activity in the form of jets and localized coronal dimmings near the coronal null point \citep{kumar2021}. These signatures occur well before (1-3 hours) the main flux-rope eruption, and are interpreted as evidence of gradual breakout reconnection that removes overlying restraining flux. Depending on the available free magnetic energy and the evolution of the flux rope, such events can produce a continuum of outcomes ranging from narrow, slow, collimated outflows to wider, faster CMEs. Although pseudostreamer CMEs have traditionally been regarded as predominantly narrow and slow \citep{wang2015,wang2018,wang2023}. Our previous results \citep{kumar2021,kumar2025} suggest that pseudostreamers can also generate more extended and energetic eruptions under favorable magnetic conditions.

In 2.5D MHD simulations, \citet{lynch2013} showed that breakout reconnection in a pseudostreamer topology can produce intermittent pre-eruption outflows and can trigger sympathetic eruptions in neighboring flux systems.
Recent 3D MHD simulations demonstrate that pseudostreamer eruptions are enabled by breakout reconnection at the coronal null, coupled with the formation and destabilization of a flux rope, producing fan-shaped CMEs that represent the large-scale extension of jet dynamics \citep{wyper2024}. \citet{lynch2025} synthesized remote-sensing and in situ observables from global MHD models, showing how the observed internal structure and fine-scale plasma features of pseudostreamer CMEs depend strongly on spacecraft trajectory and magnetic connectivity. These studies establish a unified picture in which pseudostreamer jets and CMEs arise from the same fundamental reconnection-driven magnetic evolution across multiple spatial scales.

%%%%%%%%%%%%%%%%%%%%%%%%%%%%%%%%%%%%%%%%%%%%%%%%%%%%%%%%%%%%%%
\begin{figure*}
\centering{
\includegraphics[width=12cm]{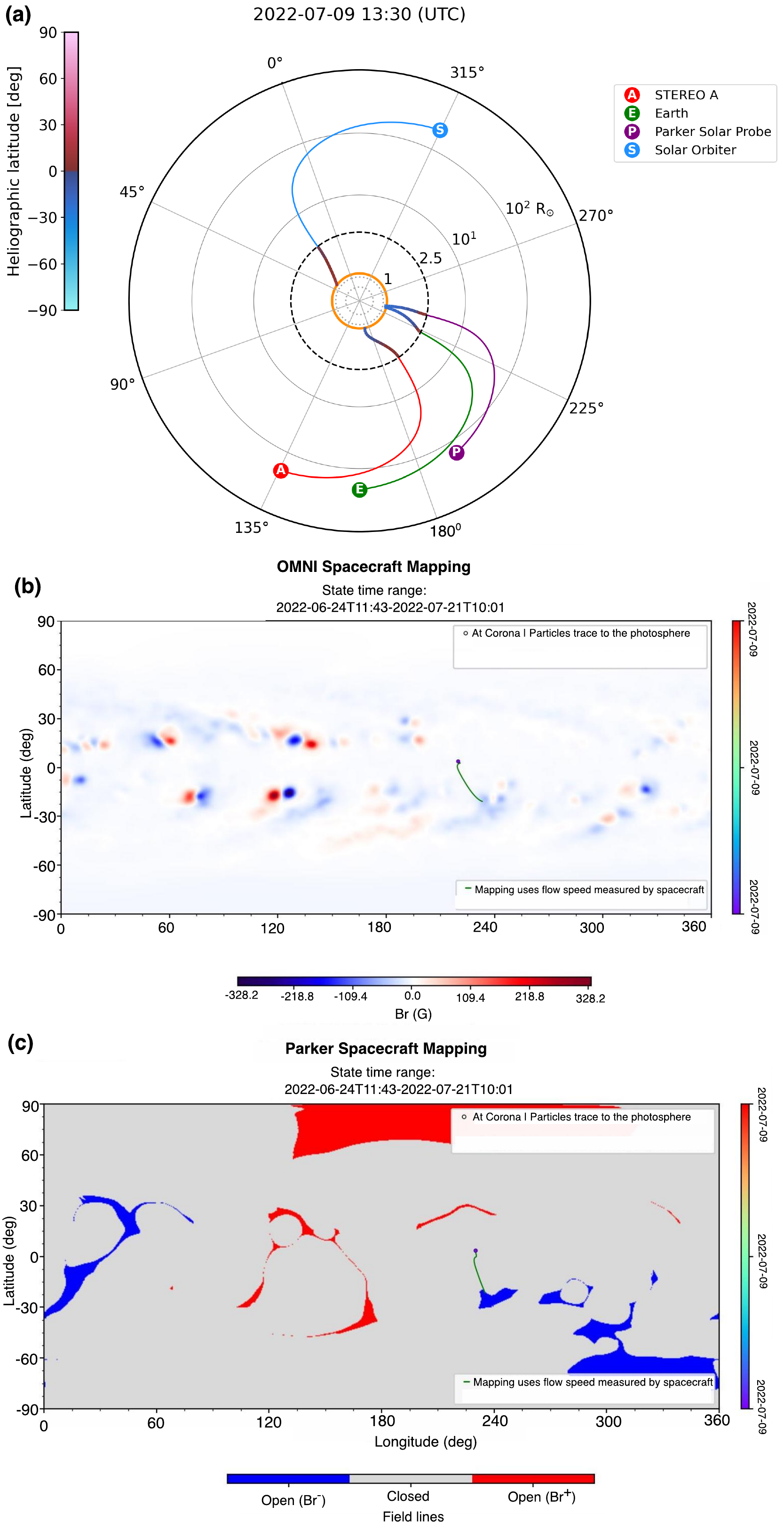}
}
\caption{Magnetic connectivity on July 9, 2022. 
(a) PFSS back-mapping results from Solar-MACH at 13:30 UT, showing the magnetic connectivity of multiple spacecraft. The color coding along the traced magnetic field lines represents the heliographic latitude of the footpoints at the solar surface. (b, c) Photospheric connectivity maps for the OMNI and PSP spacecraft using HMI CR2259 (CORHEL simulation with the Thermo-2 model) during 13:00–14:00 UT.} 
\label{app-fig2}
\end{figure*}
%%%%%%%%%%%%%%%%%%%%%%%%%%%%%%%%%%%%%%%%%%%%%%%%%%%%%%%%%%%%%%%

%para on pseudostreamer wind
Fast solar wind streams are known to originate predominantly from coronal holes, while the sources of the slow solar wind appear to be more diverse and are closely associated with large-scale coronal magnetic topologies \citep{viall2020,wang2024}. In particular, pseudostreamers have been identified as important contributors to slow and intermediate-speed solar wind streams \citep{ywang2012a,Abbo2016,kasper2021}. Both observational and modeling studies suggest that plasma release from pseudostreamers should be governed by interchange reconnection at open–closed field boundaries \citep{antiochos2011,antiochos2012,scott2018,lynch2023}. The eruptions in small-scale null-point topologies can also generate recurrent jets and localized heating \citep{pariat2010,karpen2017,wyper2017,wyper2022}, thereby supplying mass and energy flux to the heliosphere.

The implications of pseudostreamer eruptions extend beyond coronal dynamics. Fast CMEs originating from pseudostreamer regions can drive shocks in the low corona and interplanetary medium, generating Type II radio bursts and accelerating SEPs \citep{kumar2025}. Because pseudostreamers are bounded on both sides by open magnetic flux of the same polarity, they may offer favorable magnetic connectivity for particle escape during interchange reconnection. Understanding the coupling between null-point reconnection, jet activity, and fast CMEs in pseudostreamer systems is therefore essential for constraining models of CME initiation, coronal shock formation, and SEP production. The interchange reconnection in multiscale pseudostreamers provides a direct pathway for energetic particles to access open magnetic field lines \citep{masson2013,masson2019,kumar2016a,chen2018,Pallister2021}, linking these eruptions to escaping electron beams (Type III radio bursts), SEPs, and heliospheric disturbances \citep{reames1999,reames2021,gopal2006,lario2016,lario2024}.

Despite advances in understanding eruptions from coronal null-point topologies, key questions remain. In particular, the locations of particle acceleration during jets and CME onset are not well constrained, and it is still unclear where these particles are accelerated and released into the heliosphere. Our study addresses these questions by analyzing multipoint and multispacecraft observations from recent missions, providing a more complete view of jet and CME onset, particle acceleration, and particle escape. 

Here, we analyse a near-limb pseudostreamer eruption associated with a C-class flare that occurred on 2022 July 9, and was observed from multiple viewpoints by Solar Dynamics Observatory\citep[SDO;][]{lemen2012}, Solar TErrestrial RElations Observatory Ahead \citep[STEREO-A; ][]{howard2008}, and Solar Orbiter \citep[SolO; ][]{muller2020}. The eruption was triggered by slow breakout (interchange) reconnection near the null, which produced quasiperiodic jets and Type IIIs, followed by a fast CME with an accompanying shock) and associated SEPs observed by Parker Solar Probe \citep[PSP; ][]{fox2016}, Wind \citep{wilson2021}, and STEREO-A. Based on radio/EUV observations, we establish a direct connection between the Type III bursts and the quasiperiodic jets (period$\approx$5, 10 min) originating at the open–closed boundary (separatrix) of the pseudostreamer. We confirm that interchange reconnection near the null point drives the coronal rain frequently observed in such null-point topologies. We discuss the propagation of a fast EUV wave (shock) through a neighboring null-point topology and the associated MHD-wave mode conversion. 
In \S3, we present the observations, analyse key multiwavelength signatures of the underlying physical processes, and compare the results with a 3D MHD simulation of a pseudostreamer eruption. \S4 discusses the evolving roles of interchange and flare-type reconnection and associated particle acceleration in this event.  We summarize the main conclusions of this work in \S5.
%%%%%%%%%%%%%%%%%%%%%%%%%%%%%%%%%%%%%%%%%%%%%%%%%%%%%%%%%%%%%%%

%%%%%%%%%%%%%%%%%%%%%%%%%%%%%%%%%%%%%%%%%%%%%%%%%%%%%%%%%%%%%%
\begin{figure*}
\centering{
\includegraphics[width=16cm]{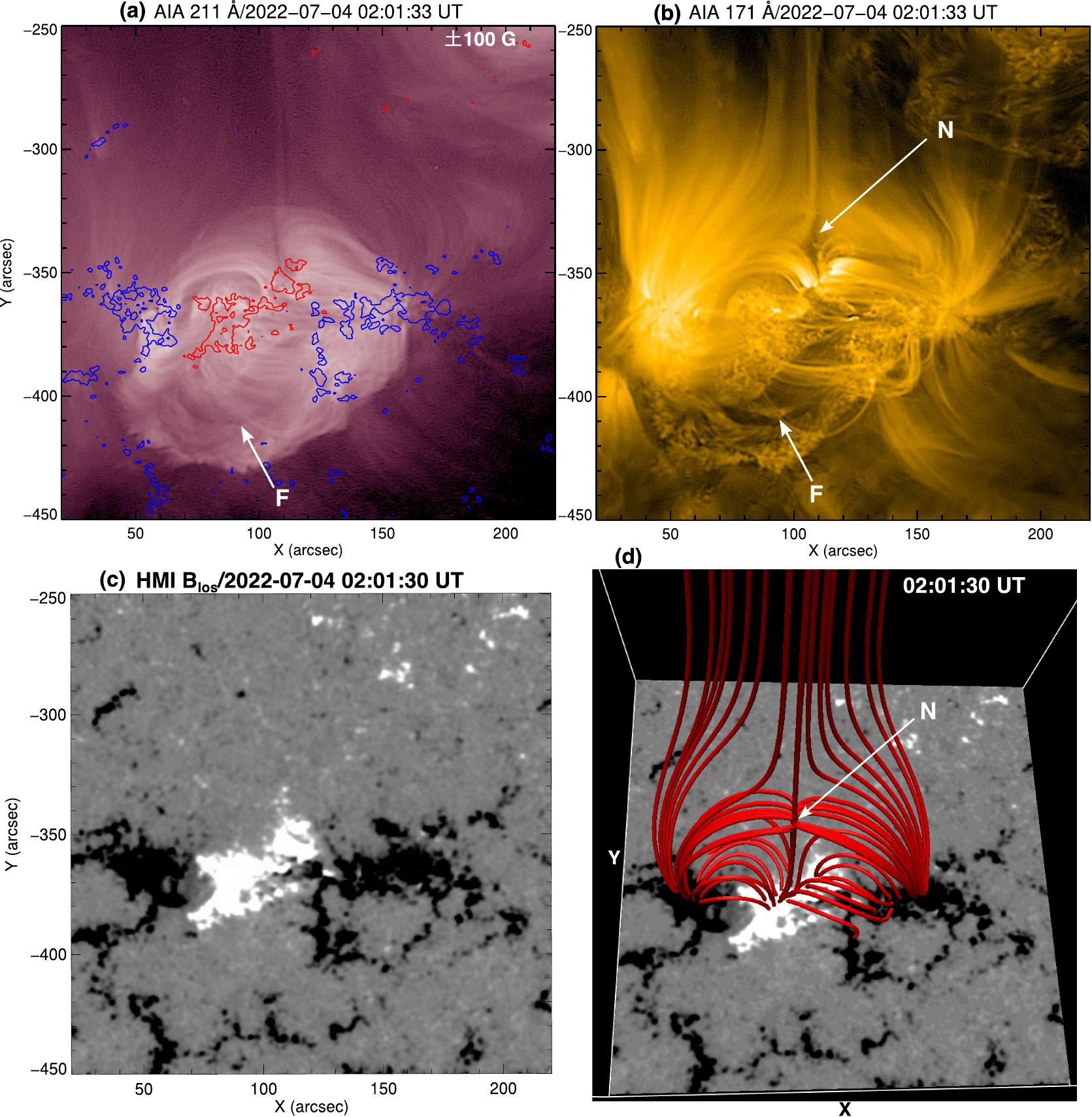}
}
\caption{Magnetic field and plasma configuration. 
(a, b) AIA 211 and 171~\AA~  images of the active region on 4 July 2022. F marks the filament. Red (blue) contours outline positive (negative) polarity regions. (c, d) HMI magnetogram (scale=$\pm$100 G) and the corresponding potential-field extrapolation of the source region at 02:01:30 UT on 4 July 2022. N denotes the null point.
} 
\label{fig-ext}
\end{figure*}
%%%%%%%%%%%%%%%%%%%%%%%%%%%%%%%%%%%%%%%%%%%%%%%%%%%%%%%%%%%%%%%

\section{Data}\label{data}
We analyzed SDO/AIA full-disk images of the Sun (field-of-view $\approx$ 1.3~$R_\sun$) with a spatial resolution of 1.5$\arcsec$ (0.6$\arcsec$~pixel$^{-1}$) and a cadence of 12~s, in the following channels: 304~\AA~ (\ion{He}{2}, temperature $T\approx 0.05$~MK), 171~\AA~ (\ion{Fe}{9}, $T\approx 0.7$~MK), and 193~\AA~ (\ion{Fe}{12}, \ion{Fe}{24}, $T\approx  1.2$~MK and $\approx 20$~MK), 131~\AA\ (\ion{Fe}{8}, \ion{Fe}{21}, \ion{Fe}{23}, i.e., 0.4, 10, 16 MK) images. The 3D noise-gating technique \citep{deforest2017} was applied to clean the SDO/AIA and SDO/Helioseimic and Magnetic Imager \citep[SDO/HMI; ][]{scherrer2012} images. 
To determine the underlying magnetic topology for jet source region, we utilized a potential-field extrapolation code \citep{nakagawa1972} available in the GX simulator package of SSWIDL \citep{nita2015}. 

The Extreme Ultraviolet Imager (EUVI; \citealt{Wuelser2004,howard2008}) aboard STEREO-A observed the studied flares occurring behind the west limb.  To track CME propagation into the interplanetary medium, we used coronagraph observations from STEREO-A/COR1 (1.3--4~$R_{\odot}$), STEREO-A/COR2 (2--15~$R_{\odot}$; \citealt{thompson2003}), and the {\it SOHO} Large Angle and Spectrometric Coronagraph (LASCO) C2 (2--6~$R_{\odot}$; \citealt{brueckner1995,yashiro2004}).

We also utilized Full Sun Imager (FSI) observations from the Extreme Ultraviolet Imager (EUI) onboard SolO, which provides full-Sun EUV images at 174 and 304~\AA~ with a cadence of about 10 minutes \citep{rochus2020}. These data offer a global view of the eruption from SolO’s vantage point.
We used X-ray data from STIX (Spectrometer/Telescope for Imaging X-rays; \citealt{krucker2020}) onboard SolO, which provides imaging spectroscopy of solar flares. The X-ray images were reconstructed with the standard CLEAN algorithm \citep{aschwanden2004}.

We analysed PSP solar wind and energetic particle observations from the Electromagnetic Fields Investigation \citep[FIELDS; ][] {bale2016}, the Solar Wind Electron Alpha Proton \citep[SWEAP; ][]{kasper2016, kasper2021}, and the Integrated Science Investigation of the Sun \citep[IS$\sun$IS; ][]{mccomas2016}. 

We analyzed radio imaging observations at metric and decimetric frequencies (150–450 MHz) from the Nançay Radioheliograph (NRH; \citealt{kerdraon1997}). Dynamic radio spectra covering metric and decimetric emissions from the low corona were obtained from the Radio Solar Telescope Network (RSTN) Learmonth Observatory and the e-Callisto network (extended Compound Astronomical Low-frequency Low-cost Instrument for Spectroscopy and Transportable Observatory; \citealt{benz2009}). Interplanetary radio emissions were provided by Wind/WAVES observations \citep{bougeret1995}.

SEP electrons and protons at 1 AU were analyzed using observations from Wind/3DP (Three-Dimensional Plasma and Energetic Particles; \citealt{lin1999}), SOHO/EPHIN (Electron Proton Helium Instrument; \citealt{muller1995}), SOHO/ERNE (Energetic and Relativistic Nuclei and Electron; \citealt{torsti1995}), and the STEREO-A/IMPACT (In situ Measurements of Particles and CME Transients) instrument suite  \citep{luhmann2008}.

Figure~\ref{app-fig2}(a) shows the heliospheric configuration of the inner heliosphere on 2022 July 9 at 13:30~UT, projected onto the heliographic equatorial plane. The positions and magnetic footpoint connectivities of STEREO-A, Earth, Parker Solar Probe (PSP), and Solar Orbiter are displayed in polar coordinates, with radial distance (in $R_{\odot}$) and longitude. Earth and PSP were separated by $\approx$32$^\circ$ in longitude, with PSP located closer to the Sun (0.74~AU), while STEREO-A was separated from Earth by $\approx$25$^\circ$ in longitude. Solar Orbiter was located near the ecliptic plane at a heliocentric distance of $\approx$1~AU and $\approx$155$^\circ$ west of Earth.
The magnetic connectivity of these spacecraft to the Sun was computed assuming a nominal Parker spiral, using the measured solar wind speeds at the respective spacecraft locations at 13:30~UT on 2022 July 9, down to a radial distance of 2.5~$R_{\odot}$. Below 2.5~$R_{\odot}$, a PFSS magnetic field configuration was employed, derived from the SDO/HMI magnetogram at 13:30~UT on 2022 July 9, as obtained from the Solar-MAC tool \citep{gieseler2023}.
Photospheric connectivity maps for the OMNI (near-Earth) and PSP spacecraft were derived using the HMI synoptic magnetogram for Carrington Rotation 2259 (Figure \ref{app-fig2}(b,c)). The maps were obtained from a CORHEL global coronal simulation \citep{lionello2009,riley2011} employing the Thermo-2 MHD model, and show the magnetic footpoints of the spacecraft at the photosphere during 13:00–14:00 UT on July 9, 2022.

Earth and PSP were magnetically connected to neighboring coronal regions closer to the erupting pseudostreamer. In contrast, STEREO-A was connected to a region separated by approximately 40--50$^\circ$ in longitude, whereas Solar Orbiter was connected to a distant photospheric region. Furthermore, the photospheric magnetic connectivity of OMNI (near-Earth) and PSP, respectively, was obtained by back-mapping in-situ plasma parcels to the solar surface using the measured solar-wind speed (Figure \ref{app-fig2}(b,c)). Both maps indicate that the spacecrafts were connected to the localized open-field near the pseudostreamer. The SEP event was observed by PSP, Wind/3DP, SOHO/ERNE, and STEREO-A.\\

\section{RESULTS}\label{obs}
\subsection{Event overview}\label{sec-firstPS}
NOAA active region 13047 was located near the west limb (S22W88, viewed from Earth) on 2022 July 9, and produced a C8.5 flare associated with a filament eruption. The flare started at 13:29 UT, peaked at 13:48 UT, and ended at 14:05 UT.

The STEREO/EUVI-A 171~\AA~ image at 04:09:30 UT on 2022 July 8, shows a fan–spine topology and a nearby giant helmet streamer (HS) located north of the active region (Figure \ref{app-fig1}(a)). In addition, large-scale pseudostreamers (PS) were observed near the outer boundaries of both polar coronal holes. The zoomed-in view of the fan–spine topology reveals a null point (N) at about 60 arcsec above the limb (Figure \ref{app-fig1}(b)). The base width of the fan–spine topology is about 100 arcsec. The outer spines and open structures extend into the middle corona and are observed in MLSO K-Cor and LASCO C2 coronagraph images (Figure \ref{app-fig1}(c)). In addition, a small closed fan–spine topology (marked by the red arrow) is also evident near the base of the giant helmet streamer (Figure \ref{app-fig1}(b)).

To understand the magnetic configuration of AR NOAA 13047, we used the HMI magnetogram at 02:01:30 UT on 2022 July 4, when the active region was on disk. The AIA 211 and 171~\AA~ images reveal a pseudostreamer with a null point (N) and a U-shaped filament (F) lying along the circular polarity inversion line (Figure \ref{fig-ext}(a,b)). The potential-field extrapolation of the source region confirms a null-point topology, consistent with the AIA observations (Figure \ref{fig-ext}(c,d)).

%%%%%%%%%%%%%%%%%%%%%%%%%%%%%%%%%%%%%%%%%%%%%%%%%%%%%%%%%%%%%%
\begin{figure*}
\centering{
\includegraphics[width=18cm]{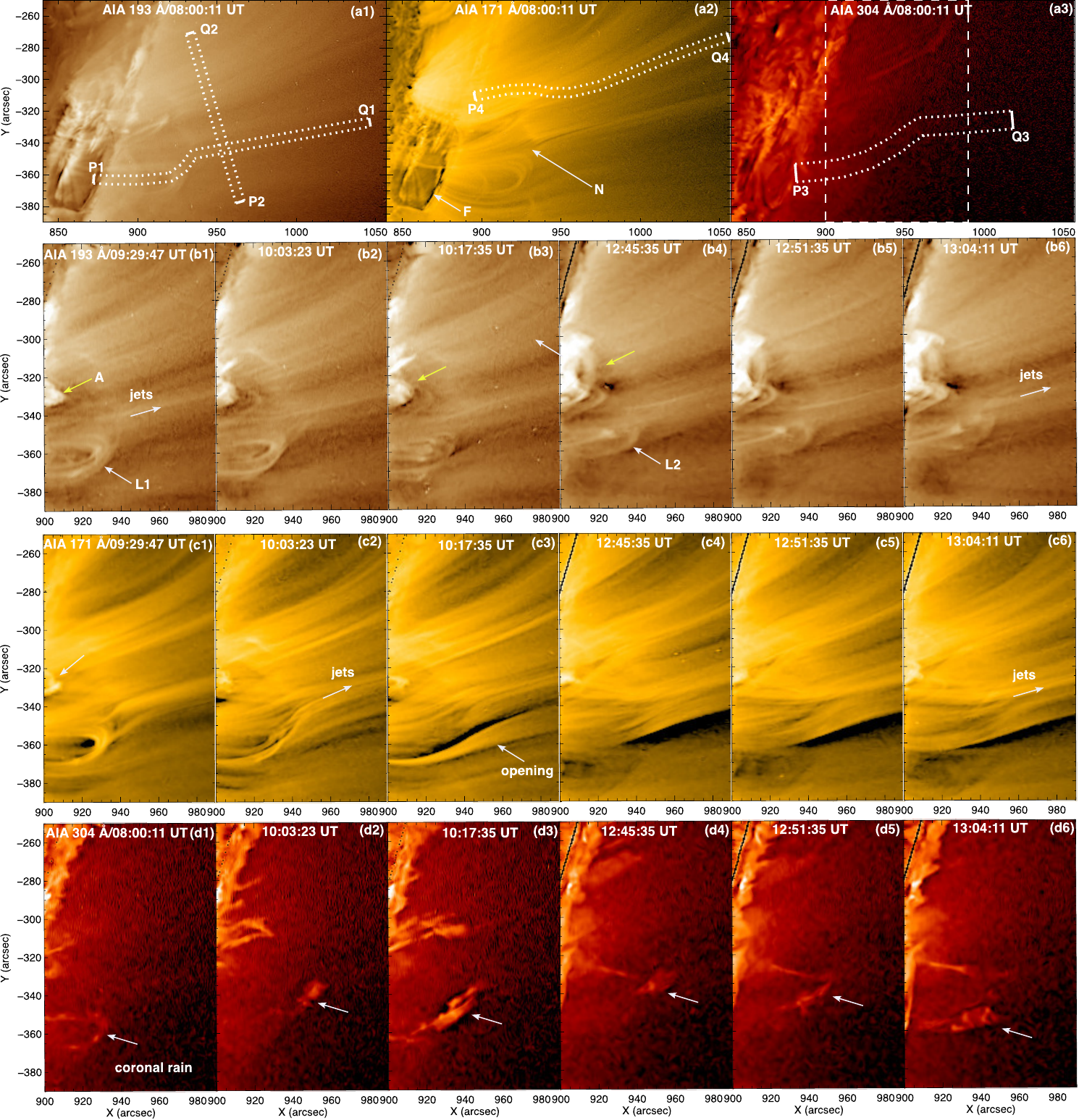}
}
\caption{Slow breakout reconnection, jets, and coronal rain prior to the pseudostreamer eruption. 
(a1--a3) AIA 193 and 171~\AA~ images showing the null-point topology. The dashed rectangle in (a3) indicates the field of view of the (b), (c), and (d) rows. N marks the approximate location of the magnetic null. P1Q1, P2Q2, and P3Q3, P4Q4 denote the slices used to construct the time–distance intensity plots shown in Figures~\ref{fig-td1} and \ref{app-fig3b}, respectively. F indicates the filament.
(b1--b6, c1--c6) AIA 193 and 171~\AA~ images illustrate the signatures of breakout reconnection associated with loop systems L1 and L2 near the null point, subsequent opening, and jets. The label A marks the arcade of hot loops formed during the pre-eruptive breakout reconnection.
(d1--d6) Formation of coronal rain (marked by white arrows) in the vicinity of the reconnection site.
} 
\label{fig-pre}
\end{figure*}
%%%%%%%%%%%%%%%%%%%%%%%%%%%%%%%%%%%%%%%%%%%%%%%%%%%%%%%%%%%%%%%

%%%%%%%%%%%%%%%%%%%%%%%%%%%%%%%%%%%%%%%%%%%%%%%%%%%%%%%%%%%%%%
\begin{figure*}
\centering{
\includegraphics[width=11.5cm]{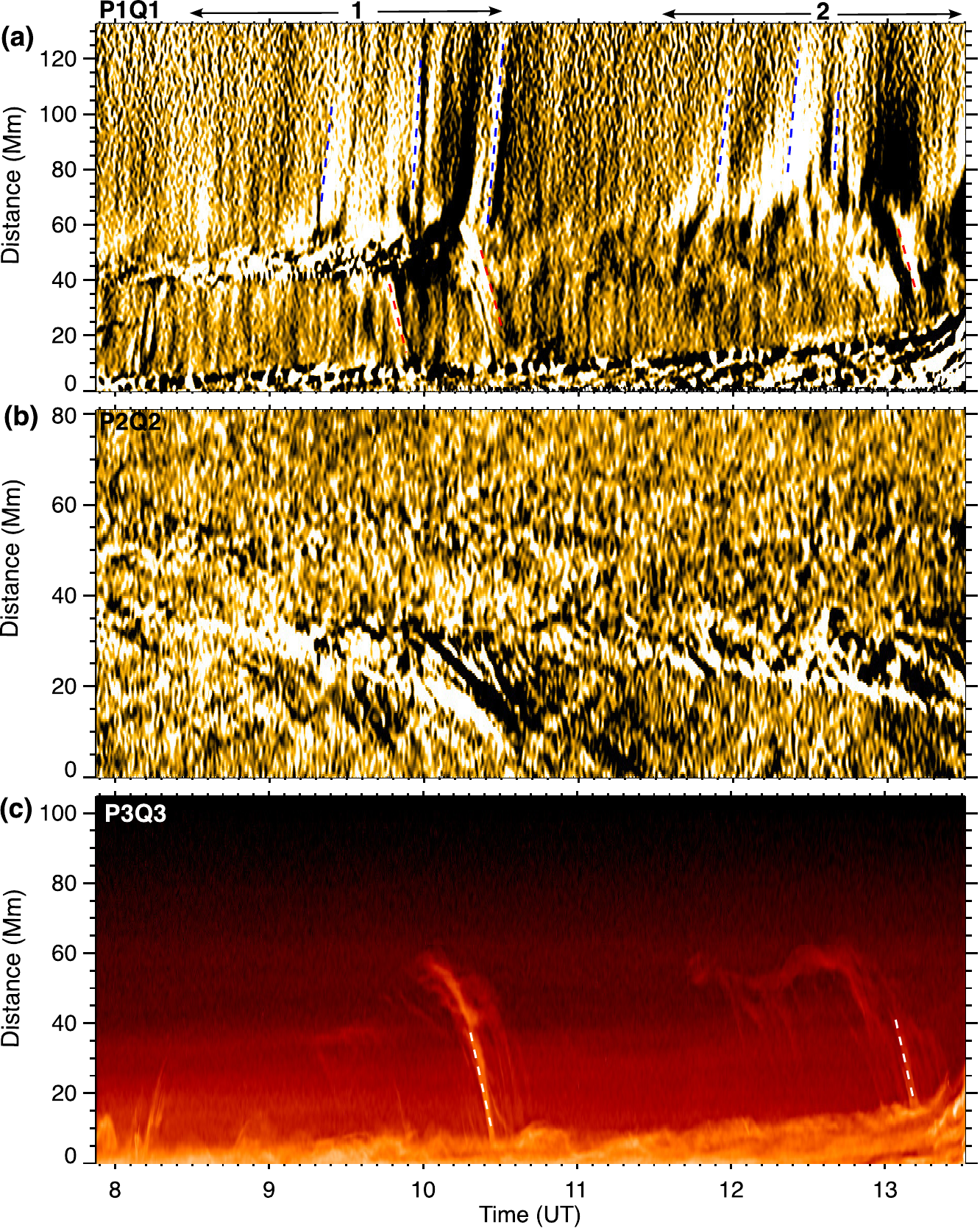}
}
\caption{Temporal evolution of jets and coronal rain during slow breakout reconnection before the pseudostreamer eruption.
(a–c) Time–distance intensity plots along slices P1Q1, P2Q2, and P3Q3 using AIA 171~\AA~(running-difference) and 304~\AA~ observations. Two episodes of recurrent jets are indicated by labels 1 and 2. The dashed lines in (a) and (c) are used to derive the speeds of the jets and coronal rain.
} 
\label{fig-jets}
\end{figure*}
%%%%%%%%%%%%%%%%%%%%%%%%%%%%%%%%%%%%%%%%%%%%%%%%%%%%%%%%%%%%%%%

\subsection{Pre-eruption signatures: Slow breakout (interchange) reconnection, faint jets, and coronal rain}
Figure~\ref{fig-pre} and associated Movie S1 shows the multiwavelength evolution of the null-point topology and filament $\approx$5.5~hr prior to the filament eruption, as observed by SDO/AIA in the 193, 171, and 304~\AA~ channels. 
During $\approx$08:30~UT and 10:30~UT, the AIA 193 and 171~\AA~ images (Figure~\ref{fig-pre}(b1--b3)) reveal the interaction of a system of coronal loops (L1) with open magnetic structures near the null point. During this interval, L1 reconnects at the open--closed boundary (separatrix), producing a sequence of recurrent jets propagating outward along the open structures. Concurrently, a dynamic arcade of hot loops (labeled $A$) forms beneath the separatrix dome during the interchange reconnection. As the jet activity continues, the L1 loop system progressively opens before the slow rise of the filament begins.

Following the first episode, a second loop system (L2) becomes visible above the filament (seen in AIA 193~\AA~ images). During $\approx$11:30~UT to 13:30~UT, L2 undergoes a similar evolution (Figure~\ref{fig-pre}(b4--b6)), interacting with the open magnetic field near the null and producing another series of jets. The repeated opening of L2 further restructures the coronal field above the filament. The flare onset occurs at $\sim$13:30~UT, marking the end of this pre-eruptive phase. The jets are best observed in the AIA 171 and 193~\AA~ channels. Jets are also detected above the open–closed boundary on the northern side of the pseudostreamer (Movie S1).
During both jet episodes, cool plasma condensations form near the reconnection region (AIA 304~\AA~ images; Figure~\ref{fig-pre}(d1--d6)) and subsequently drain along magnetic field lines as coronal rain. 

Figure~\ref{fig-jets} presents time–distance (TD) intensity plots constructed along three slits (P1Q1, P2Q2, and P3Q3 shown in Figure \ref{fig-pre}(a1,a3)) using AIA 171 and 304 channels to investigate the temporal evolution of pre-eruption dynamics. The TD intensity plot along slit P1Q1 (Figure~\ref{fig-jets}(a)) reveals recurrent jets during the pre-eruption phase. During both intervals, multiple upward propagating jets were observed. Linear fits to the jets (marked by dashed blue tracks) yield projected  speeds of 142, 256, 178, 116, 175, and 300 \kms. These jets originate near the null at the open-closed boundary ($\approx$40 Mm above the limb) and extend to heights exceeding $\approx$120 Mm. 
In addition to the outflows, returning plasma inflows are also detected along the same slit. The downflowing features, marked by red dashed tracks, occur primarily following the jet episodes around 09:50 UT, 10:25 UT, and 13:12 UT, with projected speeds of 63~\kms, 58~\kms, and 54 \kms, respectively. These downflows are co-spatial and co-temporal with the coronal rain observed in Figure~\ref{fig-jets}(c), and exhibit comparable speeds, indicating that they are associated with cooling plasma draining back along reconnected field lines.

The TD intensity plot along slit P2Q2 (Figure~\ref{fig-jets}(b)), positioned across the open structure above the magnetic null region, shows clear signatures of loop opening and lateral displacement. During the first activity interval, coronal loops exhibit gradual outward expansion followed by a systematic southward shift. A similar but more extended displacement is observed during the second episode.
Furthermore, P2Q2 captures recurrent jet signatures. Each jet appears as a localized bright enhancement with a finite transverse extent, allowing us to estimate the jet width from the full width at half maximum of the intensity profiles. We find that the jets have characteristic widths of $\approx$5–10~Mm. The repeated occurrence of these features reveals a quasiperiodic behavior, with recurrence times of $\approx$5–10~min.

The TD intensity plot along slit P3Q3 (Figure~\ref{fig-jets}(c)) captures the evolution of coronal rain. Bright descending tracks are clearly visible during both intervals, corresponding to plasma condensations draining along the field. Two prominent downflow events are observed around 9:15–10:40 UT and 11:45–13:24 UT. Linear fits to the descending tracks (white dashed lines) yield projected downflow speeds of 55~\kms and 54~\kms, respectively. The coronal rain originates from heights of $\approx$40–60 Mm (near the null/open-closed boundary) and descends toward the lower corona, indicating thermal condensation and cooling processes \citep[e.g., ][]{mason2019}.

Figure \ref{app-fig3a} presents radio signatures associated with the jet activity from the pseudostreamer between 08:00 and 13:30 UT. NRH 150 MHz radio flux density (extracted from the radio source that appeared near the pseudostreamer) exhibits pronounced temporal variability throughout the interval. Assuming a 1-fold Newkirk coronal density model, the 150~MHz plasma emission originates at a heliocentric height of $\approx$0.13 Rs above the solar surface (for fundamental emission). The metric Type III bursts are observed between 08:40–09:30 UT and again during 10:30–12:00 UT, with peak flux densities reaching about 30–35 sfu (Figure \ref{app-fig3a}(a)). The dynamic spectrum from PSP/FIELDS (1.3–19.2 MHz) shows multiple faint Type III radio bursts during the two episodes of jets associated with reconnection of L1 and L2 (Figure \ref{app-fig3a}(b)). The Wind/WAVES dynamic spectrum (0.02–13.825 MHz) reveals a storm of Type III bursts observed at higher frequencies extending coherently to lower frequencies, reaching below 1 MHz, indicating that the associated electron beams propagate into the interplanetary medium (Figure \ref{app-fig3a}(c)). The timing of the metric (150 MHz) emission enhancements and the decametric–kilometric Type III bursts closely corresponds to the episodic jet activity observed along the pseudostreamer during the same interval.

Episodic jets are observed along the slice P4Q4 (Figure~\ref{fig-pre}(a2)) between 08:00 and 13:30 UT, propagating outward along the open structures. The time–distance intensity map (Figure \ref{app-fig3b}, top panel) shows multiple recurrent ejections with varying amplitudes, indicating repeated energy release episodes in the source region.
To quantify the temporal behavior, we performed a wavelet analysis of the jet intensities extracted from the time–distance plot (between the two horizontal boundaries marked in the top panel). The wavelet power spectrum (Figure \ref{app-fig3b}, middle panel) reveals statistically significant periodicities at about 5, 10, and 30 min, above the 95$\%$ confidence level. The global wavelet spectrum confirms these dominant periods.
To examine the connection with the associated Type III radio bursts, we extracted the radio intensity from a horizontal slit at 1.5 MHz in the dynamic spectrum and carried out a similar wavelet analysis (Figure \ref{app-fig3b}, bottom panel). The radio signal exhibits the same dominant periodicities of 5, 10, and 30 min, above the 95$\%$ confidence levels. The temporal correspondence between the EUV jet activity and the Type III burst intensity variations is evident throughout the analyzed interval.
%%%%%%%%%%%%%%%%%%%%%%%%%%%%%%%%%%%%%%%%%%%%%%%%%%%%%%%%%%%%%%
\begin{figure*}
\centering{
\includegraphics[width=18cm]{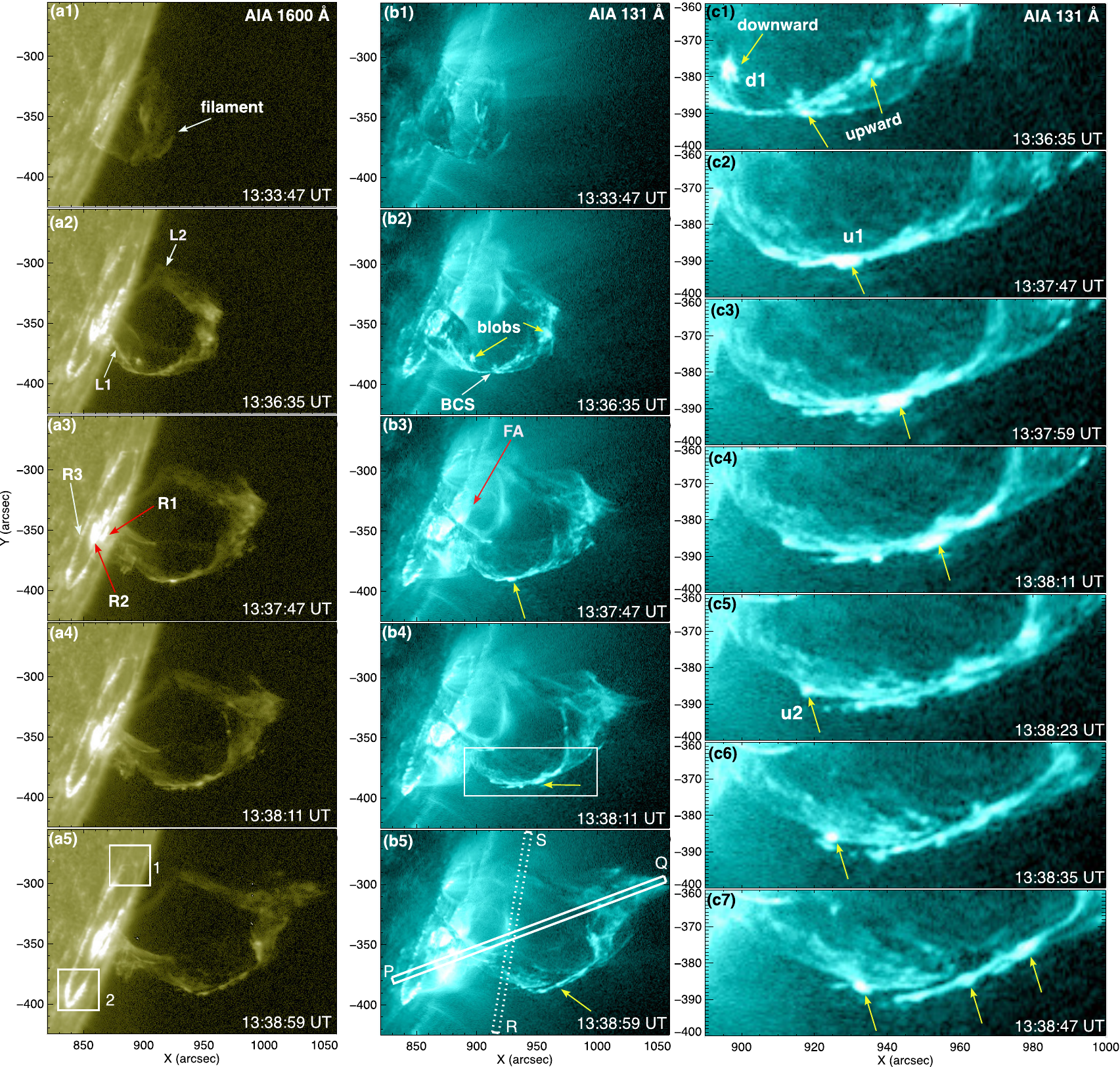}
}
\caption{Explosive breakout reconnection during the pseudostreamer eruption.
(a1–a5) SDO/AIA 1600~\AA~ images showing the erupting filament, associated flare ribbons (R1, R2), and the circular ribbon (R3). Boxes 1 and 2 indicate the regions used to extract intensity profiles shown in Figure \ref{fig-td1}. L1 and L2 are the legs of the erupting filament.
(b1–b5) AIA 131~\AA~ images showing the erupting filament and the BCS-associated plasma sheet containing multiple blobs. PQ and RS denote the slices used to construct time–distance intensity plots in FIgure \ref{fig-td1}. FA denotes the flare arcade, and BCS indicates the breakout current sheet.
(c1–c7) Enlarged views of the plasma sheet (rectangular region in panel (b4)) highlighting the evolution of multiple plasmoids (indicated by arrows).
} 
\label{fig-aia1}
\end{figure*}
%%%%%%%%%%%%%%%%%%%%%%%%%%%%%%%%%%%%%%%%%%%%%%%%%%%%%%%%%%%%%%%
%%%%%%%%%%%%%%%%%%%%%%%%%%%%%%%%%%%%%%%%%%%%%%%%%%%%%%%%%%%%%%
\begin{figure*}
\centering{
\includegraphics[width=18cm]{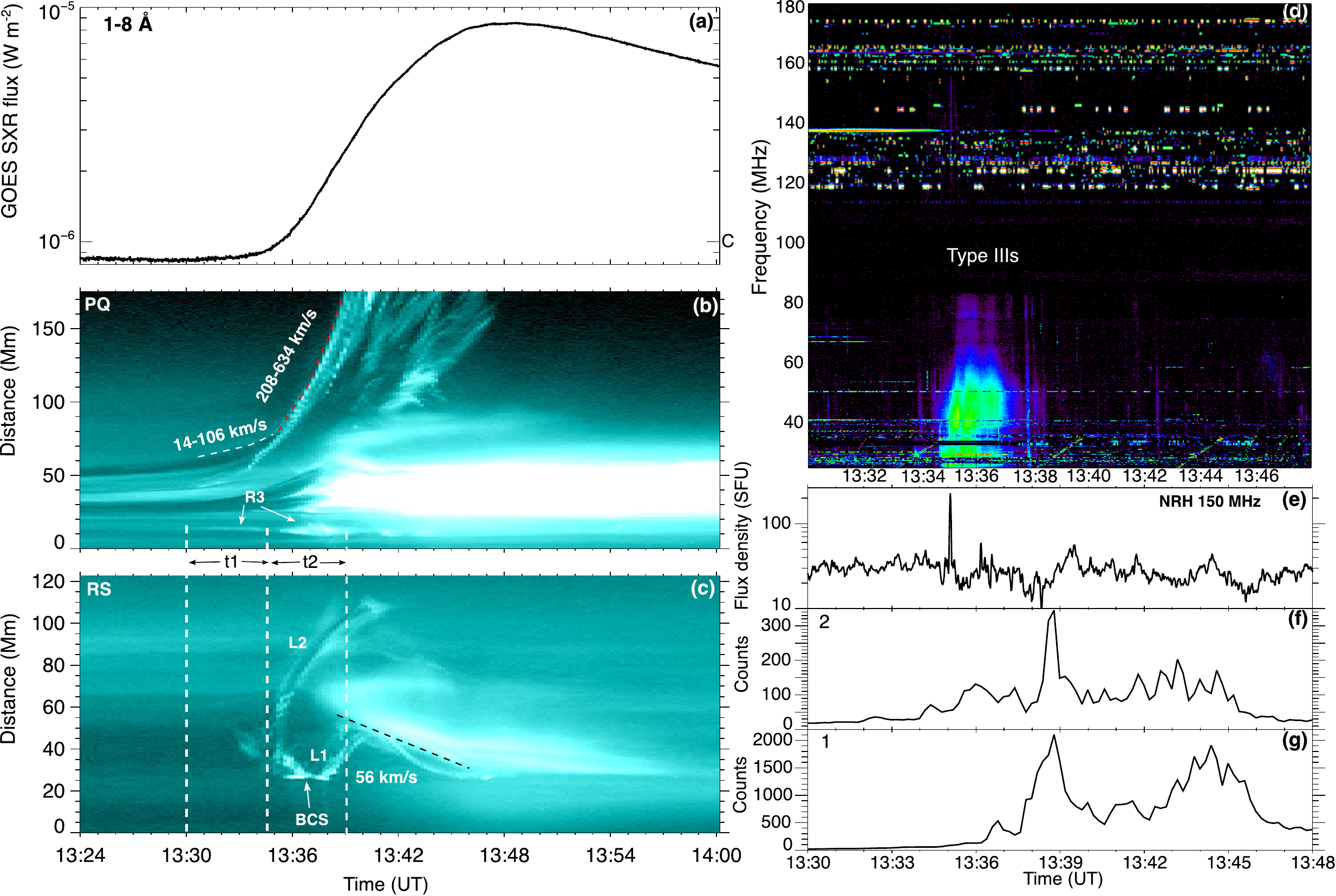}
}
\caption{Eruption kinematics associated with breakout and flare reconnection during the pseudostreamer eruption.
(a) GOES soft X-ray flux in the 1–8~\AA~ channel showing a C8.5-class flare.
(b,c) Time–distance intensity plots constructed from AIA 131~\AA~ images along slices PQ and RS (marked in Figure~\ref{fig-aia1}(b5)). The times $t_1$ and $t_2$ denote the slow and explosive phases of breakout reconnection, respectively. R3 indicates the circular ribbon, while BCS corresponds to the breakout current sheet. L1 and L2 are the legs of the erupting filament.
(d) Dynamic radio spectrum from the RSTN San Vito station (25–180 MHz). (e) NRH radio flux density profile (1 sfu = $10^{-22} W m^{-2} Hz^{-1}$) at 150~MHz.
(f,g) Temporal evolution of the intensity (peak counts, arbitrary unit) in the northern and southern portions of the circular ribbon (extracted from boxes 1 and 2 marked in Figure~\ref{fig-aia1}(a5)) using SDO/AIA 304~\AA~ images.
} 
\label{fig-td1}
\end{figure*}
%%%%%%%%%%%%%%%%%%%%%%%%%%%%%%%%%%%%%%%%%%%%%%%%%%%%%%%%%%%%%%%

\subsection{Explosive breakout reconnection}
Figure~\ref{fig-aia1}(a1--a5) and Movie S3 shows the evolution of the event in the SDO/AIA 1600~\AA~ channel, which is sensitive to emission from the upper photosphere and transition region. At 13:33:47~UT (Figure~\ref{fig-aia1}(a1)), the filament eruption is underway, with its slow rising and expanding, and localized brightenings are beginning to appear beneath it. Subsequently, during 13:35--13:37~UT, two parallel flare ribbons (R1 and R2) become clearly visible, corresponding to the footpoints of newly reconnected magnetic field lines (Figure~\ref{fig-aia1}(a2--a5)). In addition, a remote circular ribbon (R3) forms during the interaction of the erupting filament with the ambient magnetic field in the fan--spine topology. The flare ribbons continue to intensify and expand through 13:38:11~UT (Figure~\ref{fig-aia1}(a4--a5)).

Figure~\ref{fig-aia1}(b1--b5) and Movie~S2 present the corresponding evolution in the hot AIA 131~\AA~ channel. At 13:36:35~UT (Figure~\ref{fig-aia1}(b2)), a bright plasma sheet appears at the outer boundary of the southern leg of the erupting filament, which we interpret as the breakout current sheet (BCS). Multiple bright, compact intensity enhancements (``blobs") appear within the BCS and propagate both upward and downward along the current sheet. At the same time, the filament legs (L1 and L2) exhibit a clear counterclockwise kinking or rotational motion, suggestive of twist release during the eruption. At 13:37:47~UT (Figure~\ref{fig-aia1}(b3)), a hot flare arcade (FA) becomes visible beneath the erupting filament, connecting to the chromospheric ribbons R1 and R2 observed in the AIA 1600~\AA~ channel. The continued evolution of the plasma sheet and flare arcade is shown in Figure~\ref{fig-aia1}(b4--b5) during 13:38:11--13:38:59~UT.

Figure~\ref{fig-aia1}(c1--c7) illustrates a zoomed view of the BCS plasma sheet, highlighting the temporal evolution of multiple bright blobs. At 13:36:35~UT (Figure~\ref{fig-aia1}(c1)), several compact intensity enhancements are visible within the plasma sheet. These features subsequently evolve and propagate along the BCS at the outer boundary of the southern leg of the erupting filament.
Multiple blobs move upward along the plasma sheet during 13:36-13:38 UT (Figure~\ref{fig-aia1}(c2--c7)), while a few propagate downward. The characteristic size of the blobs is $\approx$2–3 arcsec. Two representative upward-moving blobs exhibit speeds of $\approx$730~km~s$^{-1}$ and $\approx$490~km~s$^{-1}$, respectively (marked by u1 and u2), while a blob moved downward at $\approx$160~km~s$^{-1}$ (marked by d1). The plasma blobs are detected in both hot and cool AIA channels. The southern leg of the filament-supporting flux rope disconnected following reconnection at the BCS (see Movie S2).

The time-distance intensity plots constructed from AIA 131~\AA~ images along slices PQ and RS (locations marked in Figure~\ref{fig-aia1}(b5)) capture the height--time evolution of the erupting filament and associated reconnection signatures. The rising speeds of the filament are derived from a second-order polynomial fit to the height–time measurements. Along slice PQ, the erupting filament rose slowly starting at $\approx$13:30~UT, with projected speeds increasing in the range $\approx$14--106~km~s$^{-1}$ (Figure~\ref{fig-td1}(b)). This slow-rise phase persists until about 13:34~UT and is accompanied by localized brightenings below the filament. The onset of these brightenings coincides with the formation of the circular ribbon (R3), indicating the initiation of slow breakout reconnection (marked by $t_{1}$). 
Following the slow-rise phase, the eruption undergoes rapid acceleration, with the speed increasing to $\approx$208--634~km~s$^{-1}$ during 13:34:30--13:39~UT, during the explosive breakout reconnection phase (marked by $t_{2}$).

The time--distance intensity plot along slice RS (Figure~\ref{fig-td1}(c)) reveals the evolution of the filament legs (L1 and L2) and the BCS. A compact brightening associated with the BCS becomes evident during the explosive breakout phase ($t_{2}$), followed by enhanced fan-loop motion toward the cusp with a speed of $\approx$56~km~s$^{-1}$. Recurrent post-eruption outflows were also detected above the cusp during 13:45--13:56~UT (Movie S2). 

%%%%%%%%%%%%%%%%%%%%%%%%%%%%%%%%%%%%%%%%%%%%%%%%%%%%%%%%%%%%%%
\begin{figure*}
\centering{
\includegraphics[width=11cm]{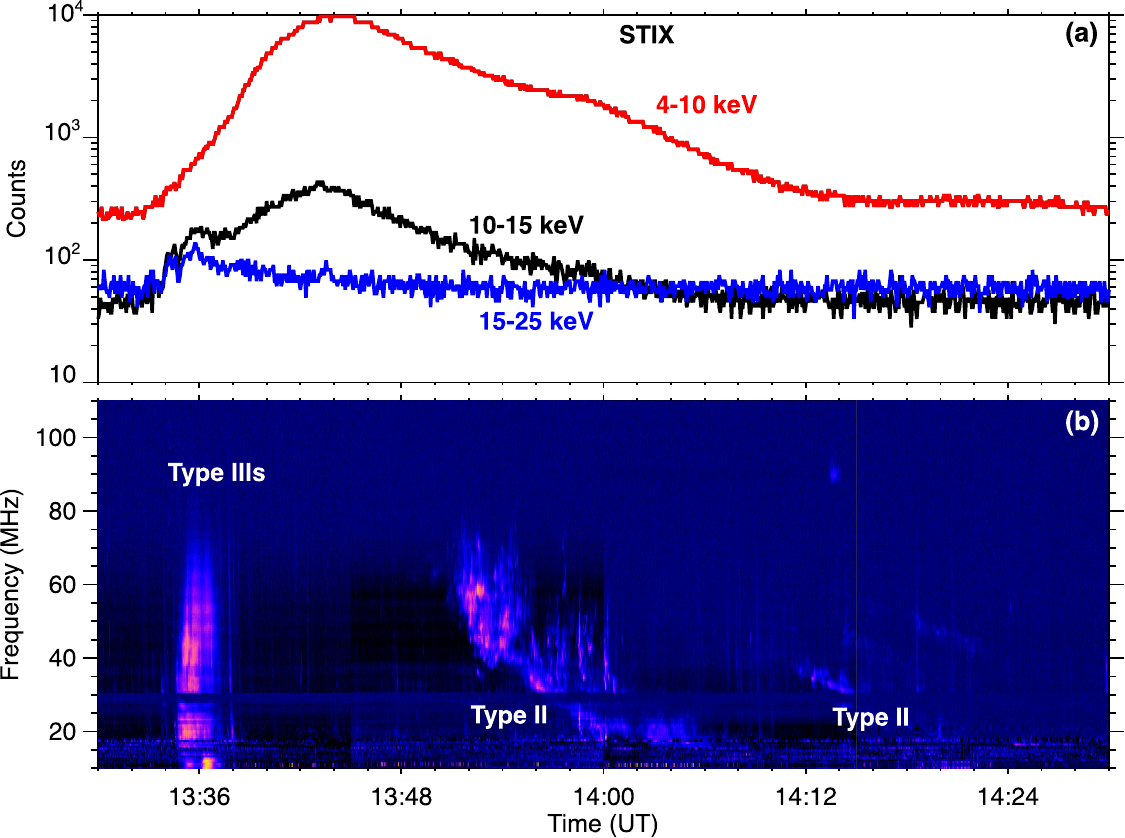}
}
\caption{X-ray fluxes and dynamic radio spectrum.
(a) X-ray count rates in different energy bands (4–10, 10–15, and 15–25 keV) measured by Solar Orbiter/STIX.
(b) Dynamic radio spectrum in the 10–110 MHz range observed by the e-Callisto Greenland station.
} 
\label{fig-flux}
\end{figure*}
%%%%%%%%%%%%%%%%%%%%%%%%%%%%%%%%%%%%%%%%%%%%%%%%%%%%%%%%%%%%%%%

%%%%%%%%%%%%%%%%%%%%%%%%%%%%%%%%%%%%%%%%%%%%%%%%%%%%%%%%%%%%%%
\begin{figure*}
\centering{
\includegraphics[width=16.5cm]{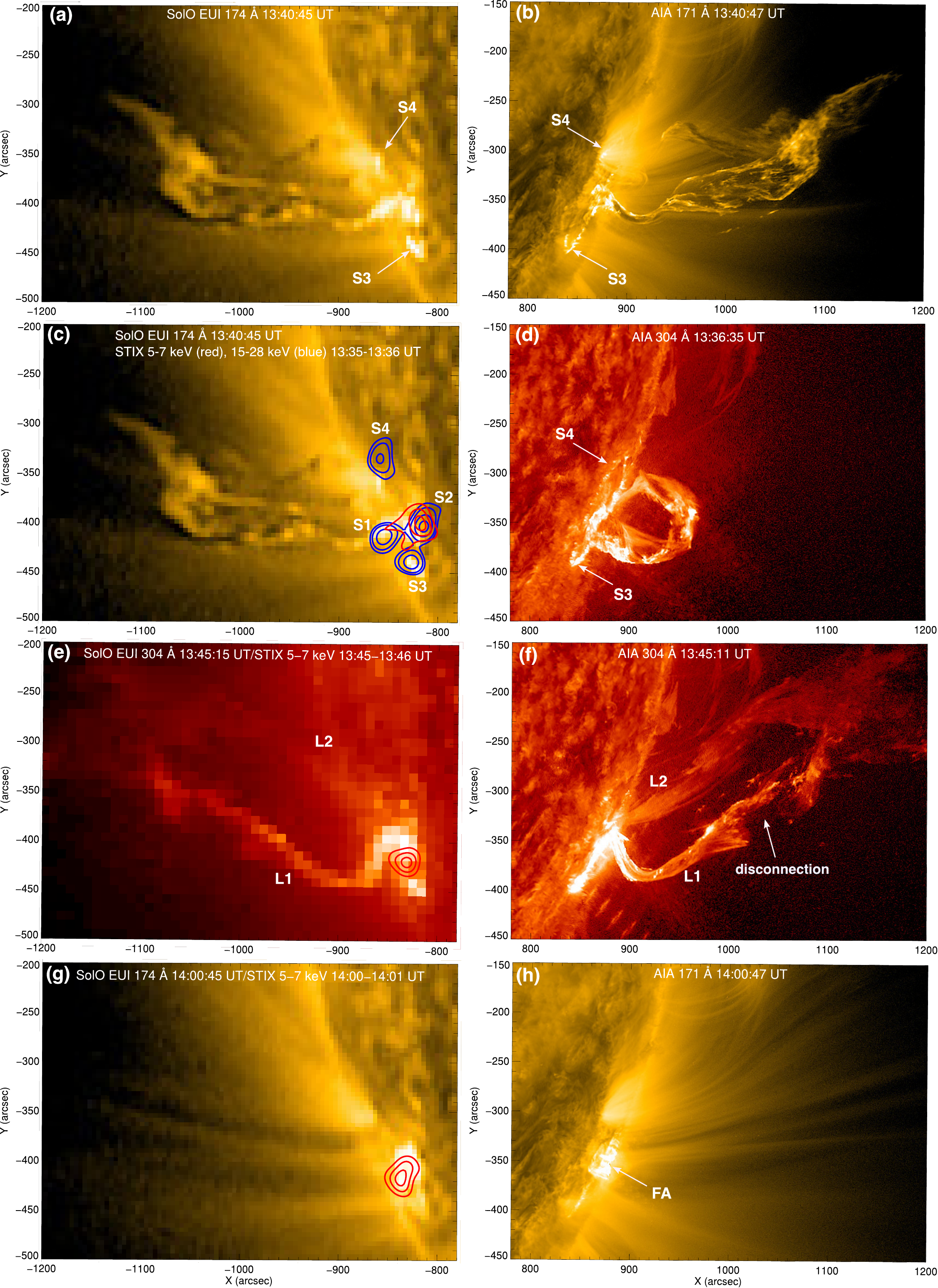}
}
\caption{Evolution of X-ray sources during the pseudostreamer eruption. 
(a, c, e, g) SolO/EUI 174 and 304~\AA~ images during the eruption, overlaid with SolO/STIX X-ray contours at 5–7 keV (red) and 15–28 keV (blue). The contour levels are 50$\%$, 70$\%$, and 90$\%$ of the peak intensity. 
(b, d, f, h) Nearly simultaneous SDO/AIA 171 and 304~\AA~ images from a different viewing angle.
L1 and L2 denote the legs of the erupting filament. S1, S2, S3, and S4 (blue contours) represent the hard X-ray footpoint sources observed during the impulsive phase of the flare. FA indicates the flare arcade.
} 
\label{fig-xray}
\end{figure*}
%%%%%%%%%%%%%%%%%%%%%%%%%%%%%%%%%%%%%%%%%%%%%%%%%%%%%%%%%%%%%%%
%%%%%%%%%%%%%%%%%%%%%%%%%%%%%%%%%%%%%%%%%%%%%%%%%%%%%%%%%%%%%%
\begin{figure}
\centering{
\includegraphics[width=9cm]{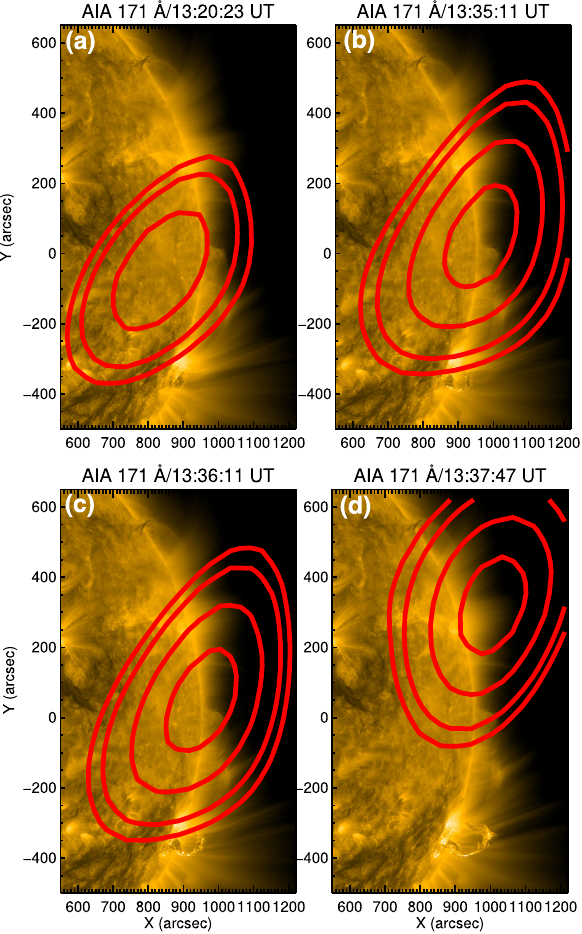}
}
\caption{Imaging of the radio source prior to and during the eruption. 
(a-d) AIA 171~\AA~ images shortly before and during the eruption, overlaid with NRH 150 MHz contours (red). The contour levels are 40$\%$, 50$\%$, 70$\%$, and 90$\%$ of the peak intensity. Panels (b), (c), and (d) show intense Type III bursts during the filament eruption. 
%(e) NRH radio flux density profile (1 sfu = 10^{-22} $W m^{-2} Hz^{-1}$) at 150~MHz.
%(f) Dynamic radio spectrum in the 10--110~MHz range observed by the e-Callisto Greenland station.
} 
\label{fig-radio}
\end{figure}
%%%%%%%%%%%%%%%%%%%%%%%%%%%%%%%%%%%%%%%%%%%%%%%%%%%%%%%%%%%%%%%

\subsection{X-ray and radio signatures}
Figure~\ref{fig-td1}(d–g) demonstrates the temporal relationship between radio signatures of escaping electron beams and the evolution of the circular ribbon during the breakout reconnection phase. A series of recurrent Type III radio bursts (metric) is observed in the dynamic radio spectrum (25–180 MHz) from the RSTN San Vito station primarily between 13:34 and 13:38 UT (Figure~\ref{fig-aia1}(a2-b2), coincident with the explosive breakout reconnection phase identified from EUV observations (Figure~\ref{fig-td1}(d)). Figure~\ref{fig-td1}(f,g) present the temporal evolution of the EUV intensity extracted from the northern (box 1) and southern (box 2) portions of the circular ribbon (indicated in Figure \ref{fig-aia1}(a5)), respectively, using AIA 304~\AA~ images. The ribbon intensities begin to rise at 13:34 UT, closely following the onset of Type III activity. 
The southern ribbon segment exhibits a sharper and stronger intensity enhancement, peaking around 13:38–13:39 UT, whereas the northern segment shows recurrent weaker fluctuations that are temporally correlated with the Type III bursts.

The radio flux density (1-s cadence) was extracted from the NRH 150 MHz source located near the eruption site. The flux profile shows an intense Type III burst at 13:35 UT (Figure~\ref{fig-td1}(e)), consistent with the RSTN dynamic spectrum. Fluctuations in the ribbon intensity are also observed during 13:39–13:48 UT. During the same interval, AIA 131 \AA~ images show ongoing outflows from the cusp near the null point, associated with flare reconnection beneath the erupting flux rope (Movie S2). Faint Type III bursts are also observed at 150 MHz and in the RSTN 25–80 MHz frequency range during 13:39–13:48 UT.

Figure \ref{fig-flux}(a) displays the X-ray light curves measured by the STIX onboard Solar Orbiter in the 4–10, 10–15, and 15–25 keV energy bands during 13:30–14:30 UT. The 4–10 keV emission (thermal) shows a clear rise starting at $\approx$13:32 UT, reaching a pronounced maximum at $\approx$13:44–13:46 UT, followed by a gradual decay through the remainder of the interval. The 10–15 keV band exhibits a temporally correlated enhancement, peaking slightly earlier, around $\approx$13:42–13:44 UT, and decaying more rapidly than the softer band. The 15–25 keV emission (nonthermal) remains comparatively weak throughout the event, showing only a modest enhancement above the background level near $\approx$13:34–13:37 UT. The dynamic radio spectrum in the 10–110 MHz frequency range observed by the Greenland station of the e-Callisto reveals a group of intense Type III radio bursts between $\approx$13:33 and 13:38 UT, spanning frequencies from 30 to 90 MHz (Figure \ref{fig-flux}(b)). Following this, a slowly drifting Type II radio burst appears at $\approx$13:48 UT, initially near $\approx$60–70 MHz and drifting down to $\approx$20–25 MHz by $\approx$14:05 UT, consistent with emission from a coronal shock moving to progressively lower plasma densities. A second episode of faint Type II emission is detected later, starting at $\approx$14:10 UT and extending until $\approx$14:18 UT, with emission confined mainly below 40 MHz. We note that the first peak in the 10–15 keV X-ray flux is nearly co-temporal with the Type III radio bursts. Although this energy range is often dominated by thermal emission, the close temporal correspondence with the 15–25 keV enhancement and the Type III bursts indicates the presence of a nonthermal component. The Type III radio bursts occur simultaneously with the hard X-ray emission (15–25 keV), indicating the prompt escape of flare-accelerated electron beams during the impulsive phase of the eruption. In contrast, the Type II radio bursts appear later, during and after the decay phase of the X-ray emission, consistent with electrons accelerated by the CME-driven shock.

We utilized STIX observations to study the evolution of X-ray sources (5–7 keV and 15–28 keV) during the eruption. In Figure \ref{fig-xray}(c), STIX X-ray contours are overplotted on near-simultaneous SolO EUI images during the flare, while AIA images provide a higher-resolution view of the eruption from the SDO  vantage point. Since EUI  (at a cadence of 10 minutes) missed the impulsive phase at 13:35–13:36 UT, X-ray contours are shown on the EUI image at 13:40 UT. To compare the X-ray source locations with the parallel flare ribbons and the circular ribbon, we used AIA 304~\AA~ images to track the filament eruption during the impulsive phase ($\approx$13:36 UT; Figure \ref{fig-xray}(d)). Interestingly, four footpoint sources (S1–S4 in 15–28 keV, blue) were detected during the impulsive phase (13:35–13:36 UT; Figure \ref{fig-xray}(c)). The 5–7 keV source represents the flare loop-top (SXR) and aligns with the flare arcade observed in EUI 174~\AA~and AIA 171~\AA~channels (Figure \ref{fig-xray}(g,h)). Sources S1 and S2 correspond to the footpoints of the flare arcade, whereas S3 and S4 coincide with some kernels of the circular ribbon (Figure \ref{fig-xray}(a,b,d)), with S3 being more intense than S4. Leg L1 rotated clockwise and heated up, followed by disconnection during the interaction (reconnection) within the fan–spine configuration. Both legs of the erupting filament (L1, L2) are visible from the two different viewing angles (Figure \ref{fig-xray}(e,f)).

NRH observations show radio emission only at 150~MHz, with no detectable Type~III bursts at higher frequencies. The 150~MHz radio flux density exhibits recurrent, short-duration enhancements between $\approx$13:33:40 and 13:38:45~UT, with the most prominent peak occurring at $\approx$13:35~UT (Figure~\ref{fig-td1}(e)). These enhancements temporally coincide with the appearance of Type~III bursts in the e-Callisto dynamic spectrum. The dynamic spectrum in the 10--110~MHz range (Figure~\ref{fig-flux}(b)) shows that the emission associated with the 150~MHz peaks extends rapidly to lower frequencies. The recurrent nature of the 150~MHz brightenings and their close temporal correspondence with the low-frequency Type~III bursts indicate multiple episodes of electron beam activity during this time interval.
A radio source in the 150-MHz band was observed near the active region prior to the filament eruption (image at 13:20:23 UT; Figure \ref{fig-radio}(a)). Intense type III bursts were detected during the interaction of the filament-carrying field with the open magnetic field near the null point (i.e., breakout reconnection) between 13:33:40 and 13:38:45 UT. During this interval, the radio source moved toward the northwest (Movie S4) and later returned to its original position (Figure \ref{fig-radio}(b–d)).

%%%%%%%%%%%%%%%%%%%%%%%%%%%%%%%%%%%%%%%%%%%%%%%%%%%%%%%%%%%%%%
\begin{figure*}
\centering{
\includegraphics[width=18cm]{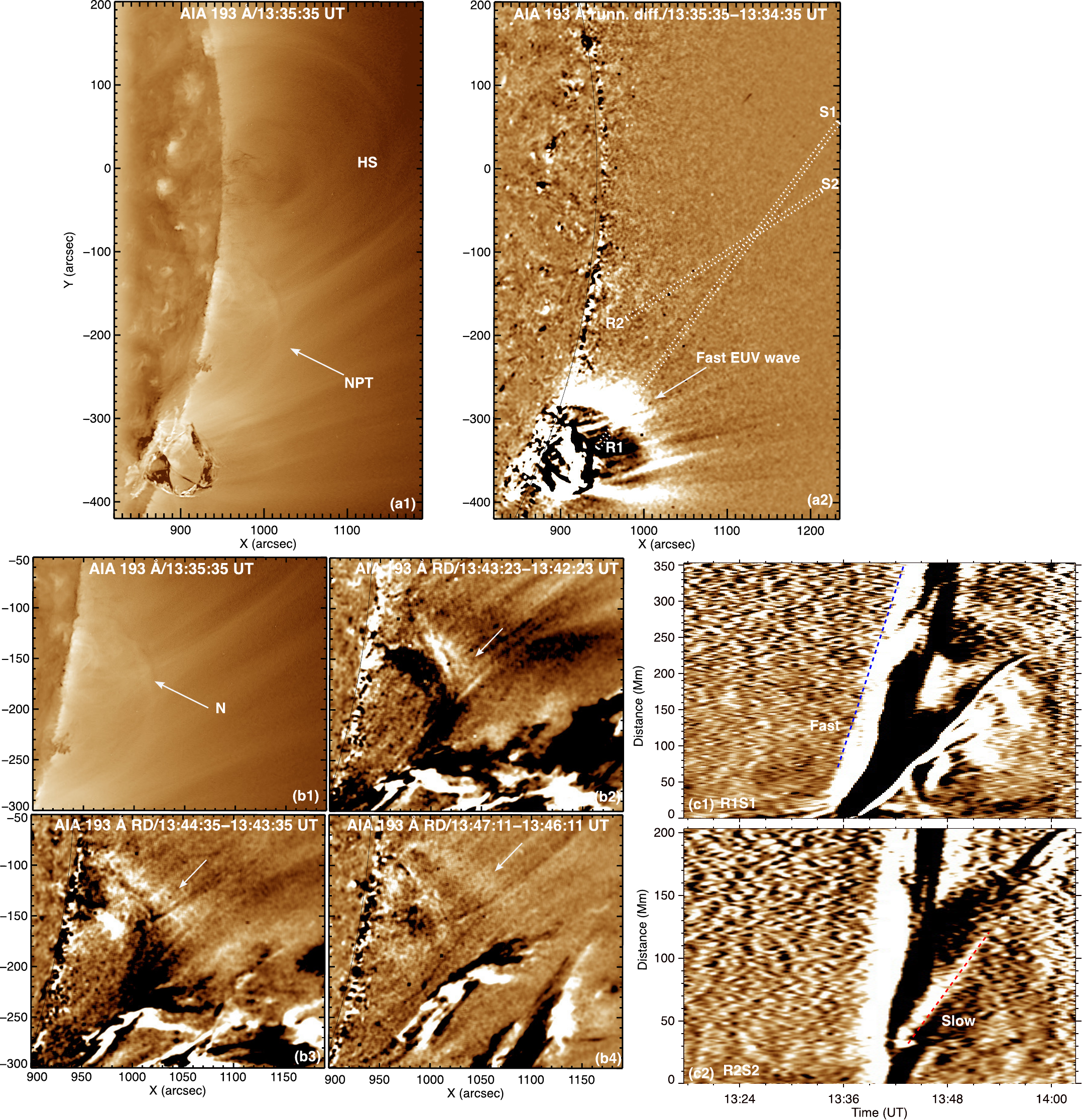}
}
\caption{Interaction of the fast EUV wave with a neighboring null-point topology. 
(a1) AIA 193~\AA~ intensity image showing the neighboring null-point topology (NPT) and the helmet streamer (HS). 
(a2) AIA 193~\AA~ running-difference image during the eruption. R1S1 and R2S2 denote the slices used to construct the time--distance maps in (c1,c2). 
(b1--b4) AIA 193~\AA~ intensity and running-difference images showing a slow-mode wavefront emerging above the null/separatrix region. 
(c1, c2) Time--distance maps along slices R1S1 and R2S2, showing the fast (610$\pm$38 \kms) and slow (158$\pm$10 \kms) wavefronts.
} 
\label{fig-mode}
\end{figure*}
%%%%%%%%%%%%%%%%%%%%%%%%%%%%%%%%%%%%%%%%%%%%%%%%%%%%%%%%%%%%%%%
\subsection{Shock interaction with nearby null-point topology and helmet streamer, MHD-wave mode conversion}
Figure~\ref{fig-mode} and associated Movie S5 shows the interaction of the fast EUV wave with the neighboring null-point topology (NPT) and the helmet streamer (HS) in AIA 193~\AA~ images. The fast EUV wave appears ahead of the erupting flux rope during the breakout reconnection phase. Between 13:36 and 14:00~UT, the fast wavefront propagates northward through the adjacent null-point topology and helmet streamer structure (Figure \ref{fig-mode}(a1,a2)). We constructed time--distance maps along slices R1S1 and R2S2 (panel a2) for the temporal evolution of wavefronts. The time--distance map along R1S1 (Figure~\ref{fig-mode}(c1)) clearly captures the fast EUV wavefront followed by the erupting filament. A linear fit yields a projected speed of $610\pm38$~km~s$^{-1}$, consistent with a fast-mode EUV wave (shock). Following the passage of the fast wave through the neighboring null region, a secondary disturbance emerged above the null/separatrix between 13:42 and 13:55~UT (Figure \ref{fig-mode}(b2--b4)). The time--distance map along slice R2S2 (Figure \ref{fig-mode}(c2)) reveals a slower propagating front originating above the null. The derived projected speed of this slow wavefront is $158\pm10$~km~s$^{-1}$, which is comparable to the coronal sound speed in a 1.2~MK plasma. 

\begin{figure*}
\centering{
\includegraphics[width=18cm]{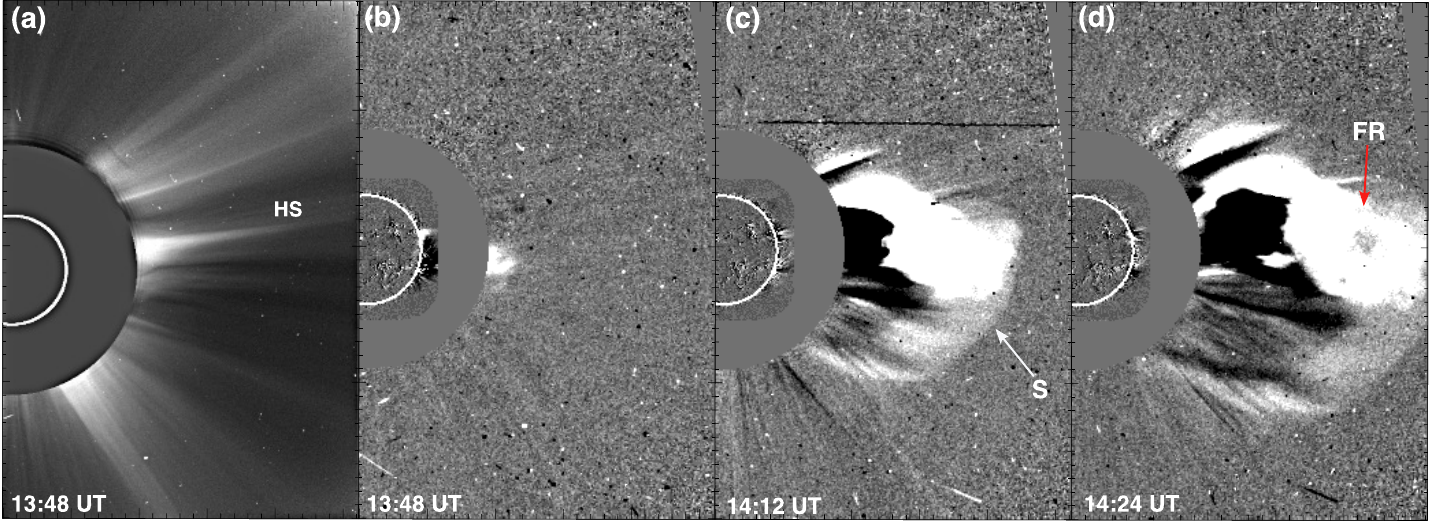}
}
\caption{SOHO/LASCO C2 coronagraph intensity (a) and running-difference (b,c,d) images showing the CME associated with the pseudostreamer eruption. HS denotes the helmet streamer, and S indicates the shock ahead of the flux rope (FR).} 
\label{fig-lasco1}
\end{figure*}

The derived speed ($\sim$600~km~s$^{-1}$) primarily represents the lateral (northward) expansion of the fast EUV wave between 13:36 and 13:43~UT. This measurement is obtained along the projected direction of propagation in the plane of the sky and therefore reflects the flank expansion of the shock. The true radial expansion speed of the shock nose is expected to be higher than the lateral component. A Type~II radio burst occurred between 13:48 and 14:00~UT, coinciding with the interaction of the shock with the dense helmet streamer in the middle corona (as seen in the LASCO/C2 coronagraph images of Figure \ref{fig-lasco1}). The shock speed derived from the Type~II frequency drift is $\sim$840~\kms, using the Newkirk (one-fold) coronal density model \citep{newkirk1961}.

The eruption produced a fast CME that first appeared in the SOHO/LASCO C2 field of view at 13:48:05 UT (Figure \ref{fig-lasco1}). The CME had a linear speed of 1034 \kms and a deceleration of -21.5 m s$^{-2}$. The CME interacted with a neighboring helmet streamer (HS). A faint shock front is visible ahead of the circular flux-rope structure. As expected, the southern leg of the flux rope appears disconnected in the coronagraph images.

\subsection{Multi-spacecraft observations of SEPs}
\begin{figure*}
\centering{
\includegraphics[width=18cm]{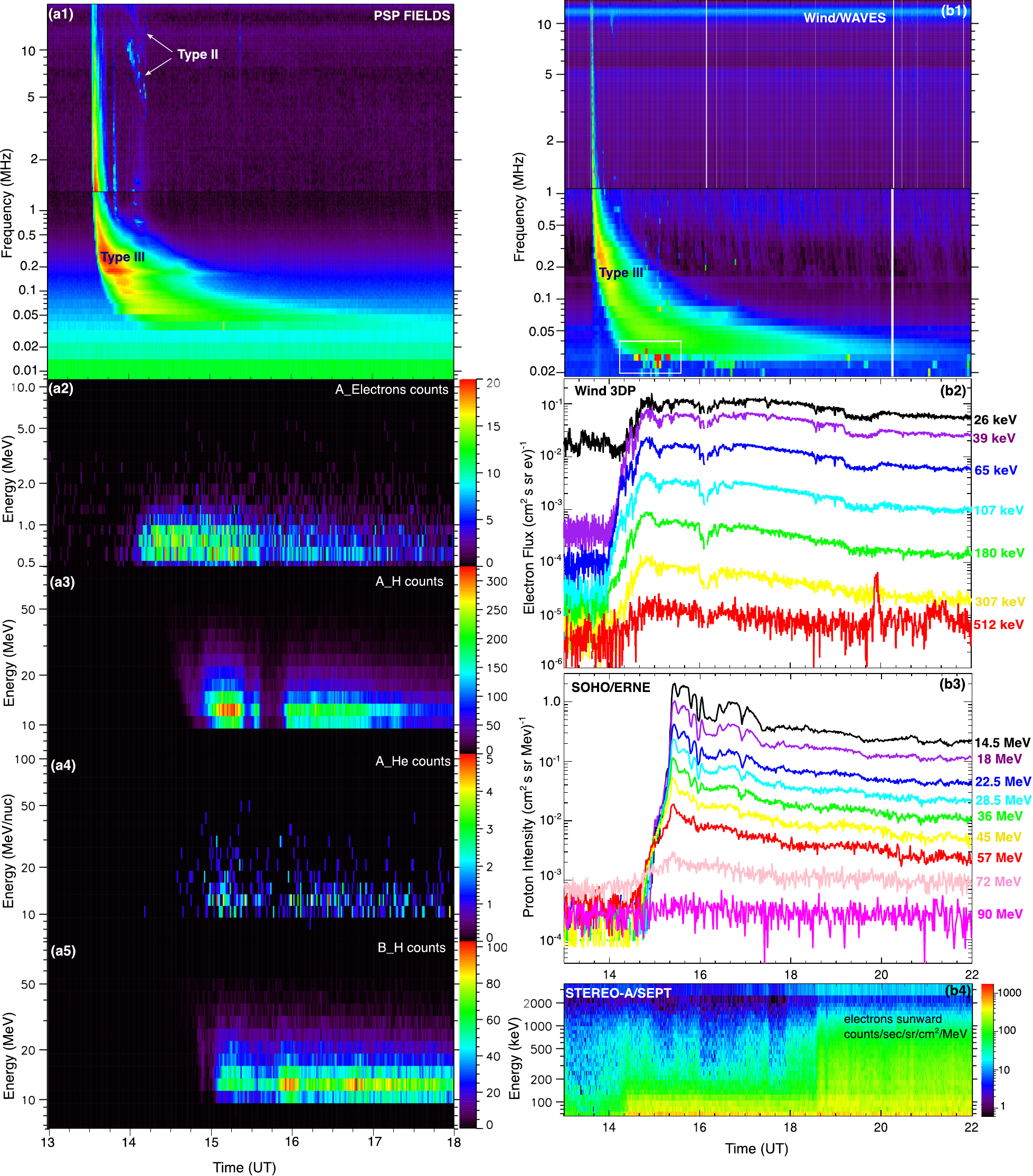}
}
\caption{SEPs detected at heliocentric distances of 0.74 and 1~AU by PSP, WIND/3DP, SOHO/ERNE, and STEREO-A/SEPT. 
(a1) Dynamic radio spectrum observed by the PSP/FIELDS instrument in the 0.01--19.2~MHz frequency range during 13-18 UT. 
(a2--a5) Electron, proton, and He count rates measured by the PSP IS$\odot$IS (EPI-Hi/HET-A and B) instrument.
(b1) WIND/WAVES dynamic radio spectrum in the 0.02--13.825~MHz frequency range during 13--22~UT. The nondrifting bright features (within the rectangular box) in the 0.02--0.035~MHz range correspond to Langmuir waves associated with the Type~III burst. 
(b2) Electron fluxes measured by WIND/3DP (12-s cadence) in the 26--512~keV energy range. 
(b3) Proton intensities observed by SOHO/ERNE (1 min cadence) in the 14.5--90~MeV energy range. 
(b4) Electrons detected by STEREO-A/SEPT (Sun-pointing, 1 min cadence) in the 65--2000~keV energy range.
} 
\label{fig-sep}
\end{figure*}

In association with the explosive breakout reconnection, a solar energetic particle (SEP) event was detected by PSP, STEREO-A, and near-Earth spacecraft such as Wind and SOHO. According to the connectivity map, SolO was not magnetically connected to the source region. No SEPs were detected at SolO. STEREO-A/WAVES dynamic radio spectrum was significantly affected by an ongoing storm of Type III bursts.
Figure \ref{fig-sep} presents a multi-spacecraft view of the SEP event as observed by PSP, Wind, SOHO, and STEREO-A, combining radio, electron, and ion measurements. The dynamic radio spectra from PSP/FIELDS and Wind/WAVES, respectively, reveal a bright Type III radio burst ($\approx$13:34-13:50 UT) extending from about 19.2 MHz to 0.02 MHz, indicative of electron beams propagating outward along open magnetic field lines (Figure \ref{fig-sep}(a1,b1)). A fainter Type II burst ($\approx$13:45-14:05 UT) is also visible at higher frequencies in the PSP FIELDS and Wind/WAVES spectra, consistent with shock-related emission near the Sun. The close agreement in onset time and frequency evolution between PSP and Wind indicates that both spacecraft observed the same radio source at the same time, despite their different heliocentric distances.

Energetic particle enhancements follow shortly after the radio signatures. PSP electron intensities (EPI-Hi/HET-A and B, Figure \ref{fig-sep}(a2–a5)) increase beginning at $\approx$14:00 UT at energies of $\approx$0.3–1 MeV, with higher-energy electrons arriving earlier than lower-energy channels, indicating clear velocity dispersion. The HET-B observations indicate that the event onset was anisotropic, with particle intensities significantly higher in HET-A than in HET-B, suggesting that the particle population was predominantly streaming antisunward. Proton and helium (He) intensities at PSP (Figure \ref{fig-sep}(a3–a5) rise later, with onsets near $\approx$14:30–15:00 UT depending on energy. 
At L1, the onset of the SEP event occurred during an out-of-the-ecliptic magnetic field configuration period. 
Wind 3DP electron fluxes (Figure \ref{fig-sep}(b2,b3)) show a rapid, nearly simultaneous increase across energies from 26 to 512 keV starting at $\approx$14:00 UT, followed by a gradual decay. The 180 keV electrons arrived at the Wind spacecraft at 13:58:41 UT. SOHO/ERNE proton intensities (Figure \ref{fig-sep}(b3)) exhibit sharp rises beginning at $\approx$15:00 UT for energies up to about 72 MeV, with progressively delayed and stronger enhancements at lower energies. The 57 MeV protons measured by SOHO/ERNE reached the spacecraft at $\approx$14:43:13 UT. STEREO-A/SEPT electrons (Sun-pointing, Figures \ref{fig-sep}(b4)) display sustained enhancements at 65-2000 keV starting near $\approx$14:00 UT, confirming the widespread longitudinal extent of the electron event. These observations demonstrate a close association between the Type III radio emission and prompt release of electrons, while the presence of an interplanetary Type II burst suggests an accompanying coronal shock that contributes to the observed electron/ion intensities. Ions may also be released during the Type III radio burst (via interchange reconnection) and arrive later due to their lower velocities compared to electrons. The non-drifting bright features (within the rectangular box, (Figure \ref{fig-sep}(b1)) in the 0.02–0.035~MHz range correspond to Langmuir waves associated with the Type~III burst \citep[e.g., ][]{thejappa2012}. The electron flux in the 26~keV energy channel (at 1~au) shows corresponding fluctuations during these intervals, suggesting a common population of energetic electrons (Figure \ref{fig-sep}(b2)).

We used velocity dispersion analysis (VDA; \citealt{krucker1999}) to estimate the solar release time and the effective interplanetary path length of SEPs. 
The method assumes that particles of different energies are released simultaneously at the Sun and propagate scatter-free along a common magnetic field line with energy-dependent velocities. 
Under these assumptions, the observed onset time $t_{\mathrm{obs}}(E)$ at a given energy $E$ is related to the particle speed $v(E)$ by
\[t_{\mathrm{obs}}(E) = t_{\mathrm{rel}} + \frac{L}{v(E)},\]
where $t_{\mathrm{rel}}$ is the particle release time at the Sun and $L$ is the effective path length traveled by the particles.
A linear fit of the observed onset times as a function of inverse particle speed yields $t_{\mathrm{rel}}$ from the intercept and $L$ from the slope.
The VDA applied to the Wind/3DP electron measurements indicates a release time of $\approx$13:37~UT and a path length of $\approx$1.6~AU (Figure \ref{app-fig4}). This estimated path length could be due to the out-of-the-ecliptic field configuration (i.e. longer than the nominal Parker spiral). 
The electron flux (Figure \ref{fig-sep}(a2)) at PSP does not exhibit a clear velocity dispersion pattern, unlike the Wind/3DP observations; therefore, the VDA method is not applied to estimate the solar release time.

We also performed the VDA using the onset times of five PSP proton energy channels covering the energy range  10.3--37.5~MeV (Figure \ref{fig-sep}(a3)). The 37.5 MeV protons detected by PSP reached the spacecraft at about 14:22$\pm$1~min UT. A linear least-squares fit of the observed onset times as a function of $1/\beta$ yielded an effective particle path length of $0.87 \pm 0.09$~AU and an inferred solar particle release time of $13{:}55{:}49 \pm 4$~min~UT. The fitted path length is consistent with the nominal Parker spiral connecting the Sun and PSP.
In contrast, the VDA of the SOHO/ERNE proton onset times (Figure \ref{fig-sep}(b3)) yielded unrealistic results (path length$\approx$0.6 AU), indicating that the assumptions of simultaneous particle release and scatter-free propagation are not fully satisfied for this event. We therefore employed a Time Shift Analysis (TSA; \citealt{krucker1999,vainio2013,gomez-Herrero2021}), assuming a conservative nominal Parker spiral path length of L=1.2-1.4 AU. Using the observed onset times of the ERNE proton channels, we infer a solar proton (18 MeV) release time of $\approx$13:54-14:03~UT.

This proton release time appears to be inconsistent with the inferred electron release time. However, an early release of protons, potentially simultaneous with the electron release, cannot be excluded. Due to uncertainties in interplanetary transport, including particle scattering and the pitch-angle distribution, the present analysis is unable to definitively rule out this possibility. In general, higher-energy electrons are expected to undergo less interplanetary scattering than the comparatively slower protons, making the electron release time estimate more reliable.

%%%%%%%%%%%%%%%%%%%%%%%%%%%%%%%%%%%%%%%%%%%%%%%%%%%%%%%%%%%%%%
%%%%%%%%%%%%%%%%%%%%%%%%%%%%%%%%%%%%%%%%%%%%%%%%%%%%%%%%%%%%%%%
%%%%%%%%%%%%%%%%%%%%%%%%%%%%%%%%%%%%%%%%%%%%%%%%%%%%%%%%%%%%%%
%%%%%%%%%%%%%%%%%%%%%%%%%%%%%%%%%%%%%%%%%%%%%%%%%%%%%%%%%%%%%%%

\subsection{Comparison with a 3D MHD simulation}
We compare our observations with recent three-dimensional MHD simulations of pseudostreamer eruptions \citep[e.g.,][]{wyper2024, lynch2025}, which provide a physical framework for interpreting the observed CME initiation and ribbon morphology. Figure~\ref{fig-sim} and the associated animation (Movie S6) reveal the eruption process in a null-point topology as captured by global MHD simulations performed in spherical geometry using the three-dimensional Adaptively Refined Magnetohydrodynamics Solver (ARMS; \citealt{devore2008}). For this simulation, ARMS solves the full set of time-dependent MHD equations starting from a multipolar coronal magnetic field configuration that naturally contains a magnetic null point. The lower boundary is line-tied at the photosphere and subjected to slow, quasistatic, time-dependent driving through prescribed surface flows narrowly centered on the pseudostreamer polarity inversion line, injecting free magnetic energy into the system. As the system evolves, this boundary driving leads to the gradual formation and thinning of a breakout current sheet near the coronal null, where reconnection begins and becomes plasmoid-mediated, progressively removing overlying restraining flux and enabling the formation of a second current sheet that creates a rising, expanding flux rope through flare reconnection. In these simulations, pre-eruption jets originate near the null, followed by the development and eruption of a flux rope; the southern leg of the flux rope reconnects through the breakout current sheet and subsequently disconnects \citep{wyper2024}. Explosive flare reconnection beneath the erupting flux rope produces two flare ribbons (R1, R2) and an associated flare arcade, while simultaneous explosive breakout reconnection generates the circular ribbon (R3). 

Figure~\ref{fig-sim}(a) shows the open-closed boundary superposed on a synthetic EUV image viewed from above at a selected time during the pseudostreamer eruption, highlighting the connection between magnetic connectivity and ribbon morphology. In Figure~\ref{fig-sim}(b), high-Q (squashing factor) regions outline separatrix and quasi-separatrix layers, open-flux regions are indicated by yellow shading, and the grayscale structure reflects the evolving complexity of the connectivity mapping at the same time as (a). During the early phase of the eruption, the Q-map is dominated by a smooth separatrix boundary that evolves slowly (Movie S6), but as the eruption progresses, increasing small-scale whirls and convoluted substructures develop along this boundary, signaling plasmoid-dominated reconnection and enhanced mixing of magnetic connectivity \citep{wyper2016,wyper2021b}. This complexity peaks during the rapid opening of the closed pseudostreamer flux and release of the CME, after which the system gradually relaxes and the original pseudostreamer topology is largely restored. The synthetic EUV image shown in (Figure~\ref{fig-sim}(c)) captures CME-related core dimming regions, fainter outlying dimming associated with interchange reconnection–driven jets/outflows, multiple plasmoids, and the development of the eruptive flare arcade (below the FCS), collectively demonstrating a direct link between the evolving magnetic connectivity in the pseudostreamer and the observed ribbon and dimming signatures (Figure~\ref{fig-sim}(c)).

Although the plasmoid dynamics shown in the simulation are resolution dependent, previous studies (e.g., \citealt{uzdensky2010,huang2012,lynch2016}) indicate that higher-resolution simulations would retain similarly large plasmoids while resolving additional small-scale plasmoids/flux ropes around them. These smaller structures would likely interact and merge with the larger plasmoids. MHD simulations have shown enhanced plasma density within these small-scale flux-rope (plasmoid) structures \citep{lynch2016}. In the present isothermal simulation, where the EUV emission is proportional to the square of the plasma density, these structures would appear as bright blobs. An important question that should be addressed in future simulation studies is how, and to what extent, such small-scale features appear in synthetic EUV emission.

 Magnetic islands/flux ropes grow by continued reconnection as they traverse the current sheet; otherwise, they would not speed up the reconnection. If plasmoids outrun those ahead of them or are constrained by periodic or closed boundaries in 2D, mergers clearly occur (e.g., \citealt{tenerani2016,huang2017}). In 3D high-resolution MHD or PIC simulations of current sheet reconnection, oblique tearing modes generally appear away from the central current sheet and the resulting plasmoids interact and merge across the sheet \citep{daldorff2022,dong2022}. We believe that we see such oblique modes in our ultra-high resolution 3D simulations of eruptive flares (Dahlin et al. 2026, in preparation), but more work is needed before we can definitively state whether 3D plasmoids in the high Lundquist regime of the Sun merge or not.

Dynamic bright blobs have been detected in images of the heliospheric plasma sheet \citep{wu2026} and in the breakout and flare plasma sheets in both eruptive flares and coronal jets (e.g., \citealt{takasao2012,kumar2013blob,liu2013,kumar2018,kumar2019b, kumar2023a,kumar2025b,patel2020,wyper2024,dahlin2025}). In all these cases, they have been interpreted as plasmoids – magnetic islands or flux ropes in 3D. Plasmoids also have been directly identified in the HCS (heliospheric current sheet) with in situ data (e.g., \citealt{reville2022, lee2026, lewis2026, wu2026}), and have been proposed as a likely explanation for the periodic density structures observed in coronagraph data \citep{viall2015}. Consequently, we suggest that plasmoids are the simplest and most likely explanation for the observed blobs.

As detailed in Sections 3.1-3.3, the EUV observations of the analyzed events reveal pre-eruption jets driven by interchange reconnection, followed by slow and explosive breakout reconnection phases at the BCS, accompanied by plasmoids and the disconnection of the southern leg of the flux rope during reconnection through the BCS. These observations are largely consistent with recent 3D MHD simulations of pseudostreamer CMEs \citep{wyper2024, lynch2025}.

\section{DISCUSSION}\label{discussion}

\subsection{Physical Interpretation of Pre-eruption Dynamics}
The multiwavelength observations of two pseudostreamer eruptions presented here provide compelling evidence that the filament eruption was preceded by prolonged magnetic restructuring driven by slow breakout reconnection near the null. The combination of recurrent jets, gradual loop opening, and sustained coronal rain suggests that the system underwent progressive magnetic flux removal and thermodynamic evolution over several hours prior to the eruption.

The interaction of the closed loop systems (L1 and L2) with the open magnetic structures near the null is consistent with interchange reconnection occurring at the open–closed separatrix boundary. The recurrent jets, with projected estimated speeds ranging from $\approx$116 to 300 km s$^{-1}$, are characteristic signatures of reconnection-driven outflows consistent with earlier pseudostreamer jet studies \citep{kumar2021}. The jet initiation heights are close to the inferred null heights.

Such behavior is consistent with the breakout reconnection scenario \citep{antiochos1998,antiochos1999,karpen2012}, in which reconnection at a coronal null removes restraining overlying flux and facilitates the gradual expansion of the filament-supporting field. The presence of two temporally separated jet episodes indicates that the breakout process proceeded intermittently, implying that the overlying magnetic field was progressively weakened through successive reconnection events. Similar pre-eruption breakout signatures have been reported in previous observations of jets and CMEs in smaller and nested null-point topologies \citep{kumar2018,kumar2021,karpen2024,kumar2024}, where repeated weak reconnection episodes gradually destabilize the magnetic system prior to eruption. The MHD simulations also yield jets, puffs, and associated dimming shortly before the eruption \citep{lynch2013,wyper2021,wyper2024,lynch2025}.

 The observed loop opening and systematic southward displacement indicate large-scale magnetic reconfiguration and redistribution of magnetic tension forces. The more extended displacement during the second episode suggests that the magnetic system became increasingly stressed and topologically simplified, ultimately enabling the filament to enter a slow-rise phase before flare onset.
The observed downward flows following several jet events correspond to plasma draining along reconnected loops after the reconnection outflows. These flows indicate the presence of newly altered magnetic connectivity linking different coronal domains, and further support the interpretation of sustained interchange reconnection.

The formation of coronal rain during both jet episodes demonstrates the thermodynamic consequences of reconnection-driven restructuring. The coronal rain originates near the reconnection region and descends along closed magnetic structures with a speed of $\approx$55 km s$^{-1}$.
Interchange reconnection can rapidly evaporate plasma within newly formed loops, followed by enhanced radiative cooling and thermal instability as the loops evolve. The observed rain therefore likely results from condensation processes triggered by reconnection-driven energy deposition and subsequent cooling. 
Coronal rain is often observed in null-point topologies \citep{mason2019,kumar2019b,kumar2021}. The observations reported here confirm that interchange reconnection heats plasma in newly reconnected flux tubes, produces jets/outflows, and leads to coronal rain through the subsequent evaporation and cooling of localized plasma.

The combined observations suggest a scenario in which slow breakout reconnection progressively removes overlying magnetic flux while simultaneously altering the plasma thermodynamics within the reconnecting loop system. The first reconnection episode involving L1 initiates gradual expansion of the overlying field and triggers the onset of filament slow rise. The second episode involving L2 further weakens the confining magnetic tension and promotes additional restructuring of the null-point dome. The prolonged duration ($\approx$5.5 hr) of these pre-eruptive signatures indicates that the system evolved through a quasistatic sequence of reconnection-driven reconfigurations rather than a single impulsive trigger. 

The NRH radio imaging and flux density profiles at 150 MHz suggest that the analyzed pseudostreamer was the source of the Type III storm during our observations, spanning several hours before and during the eruption. We observe prolonged, recurrent jet activity during the pre-eruption phase. The frequency of these jets, along with the metric and IP Type III bursts, is consistent with a common origin in the pseudostreamer region.
The simultaneous occurrence of enhanced 150 MHz emission and broadband Type III bursts indicates that the jet activity is accompanied by repeated acceleration of electron beams. The metric emission originates from the low corona, while the extension of the bursts to decametric and kilometric frequencies implies that the accelerated electrons escape along large-scale open magnetic structures into the heliosphere.
%%%%%%%%%%%%%%%%%%%%%%%%%%%%%%%%%%%%%%%%%%%%%%%%%%%%%%%%%%%%%%
\begin{figure*}
\centering{
\includegraphics[width=14cm]{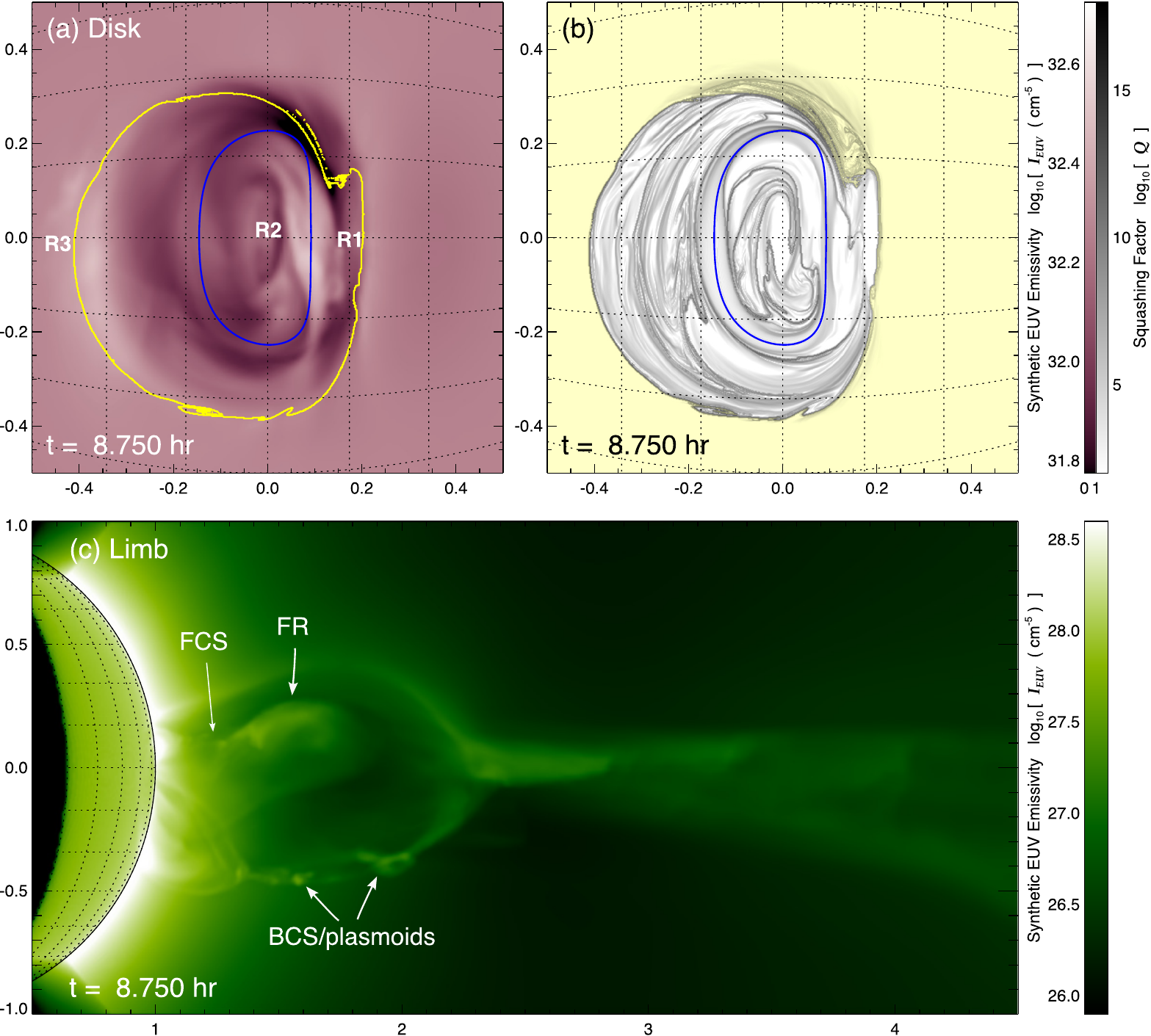}
}
\caption{MHD simulation of CME initiation in a pseudostreamer/null-point topology.
(a) Synthetic EUV emission showing the disk view of the flare ribbons (R1, R2) and the circular ribbon (R3). The yellow line marks the open–closed boundary.
(b) The B$_R$ = 0 contour (dark blue) is the polarity inversion line overlaid on log Q (a measure of magnetic field complexity). Open-flux regions are shaded in light yellow.
(c) Limb view of the pseudostreamer eruption showing synthetic EUV emission. FR denotes the erupting flux rope. BCS and FCS indicate the breakout and flare current sheets, respectively, while white arrows point out multiple plasmoids (From \citealt{lynch2025}).
} 
\label{fig-sim}
\end{figure*}
%%%%%%%%%%%%%%%%%%%%%%%%%%%%%%%%%%%%%%%%%%%%%%%%%%%%%%%%%%%%%%%

The presence of common periodicities (5, 10, 30 min) in both the EUV jets and the Type III radio bursts strongly suggests a common origin. Since Type III bursts are signatures of electron beams propagating along magnetic field lines, the matching periodicities indicate that the episodic jets and the electron acceleration events are driven by repetitive interchange reconnection at the pseudostreamer open-closed boundary.

The detection of dominant 5 and 10 min periodicities in the jet intensity suggests a possible role of slow-mode magnetoacoustic waves \citep{chen2006} in modulating the reconnection process at the pseudostreamer. Slow-mode waves, which can propagate along open structures from the lower atmosphere into the corona, may periodically perturb the magnetic configuration near the separatrix. Such compressive perturbations can periodically alter the magnetic pressure balance near the open--closed boundary, and trigger quasiperiodic interchange reconnection.
In this scenario, the observed episodic jets and associated Type III bursts represent products of intermittent reconnection driven by slow-mode wave dynamics. This mechanism requires validation through dedicated 3D MHD simulations.
 
 Interestingly, similar periodicities were reported previously in jetlets and energy release at the base of plumes \citep{kumar2022,uritsky2021,uritsky2023} and in solar wind microstreams/switchbacks observed in situ by PSP \citep{kumar2023b}. Recently, high-resolution {\it Goode Solar Telescope} (GST) observations have also imaged prolonged quasiperiodic reconnection and recurrent jets originating from the BCS formed near a coronal null point, exhibiting comparable periodicities \citep{kumar2024}. Similar periodicities are observed in null-point topologies spanning multiple spatial scales. The consistency of these timescales across different-scale magnetic configurations suggests a common underlying physical mechanism. Quasiperiodic interchange reconnection at pseudostreamers produces episodic plasma outflows and accelerated electron beams, these jets may evolve into structured solar wind streams at larger heliocentric distances. The repetitive nature of the reconnection may imprint velocity and magnetic-field fluctuations that later manifest as microstreams and switchback-like structures in the solar wind \citep{wyper2022,wyper2026}.

\subsection{Explosive breakout/flare reconnection and associated particle acceleration}
The hot plasma sheet observed near the coronal null and along the outer boundary of the southern leg of the erupting flux rope is interpreted as a breakout current sheet (BCS) forming between the closed flux of the pseudostreamer and the surrounding open magnetic field. The bright plasma sheet, along with the presence of multiple bright, compact blobs, likely indicates plasmoid instability within the reconnecting BCS.  Explosive breakout reconnection at the BCS leads to the formation of a large-scale circular ribbon (R3), while simultaneous flare reconnection below the erupting flux rope produces a typical two-ribbon flare (R1 and R2). The filament legs clearly rotate counterclockwise, suggesting the release of magnetic twist during the eruption. As the event progresses, a hot flare arcade forms and grows beneath the erupting flux rope, reflecting ongoing flare reconnection. The plasmoids travel bidirectionally along the plasma sheet, consistent with reconnection outflows, with a predominance of upward-propagating features. Their characteristic sizes ($\approx$2–3 arcsec) and measured speeds (490-730 \kms) fall within the range expected for bursty, plasmoid-mediated reconnection \citep{karpen2012,lynch2013} in a hot plasma sheet (i.e., BCS), supporting a scenario in which breakout reconnection plays a key role in both the eruption dynamics and the energization of energetic electrons. The sizes of the plasmoids are consistent with previous observations of plasmoids in flare and breakout current sheets in null-point topologies \citep{kumar2018,kumar2019b,kumar2023a,kumar2024,karpen2024,kumar2025}. The slow rise of the filament is associated with slow breakout and flare reconnection, whereas the explosive breakout and subsequent flare reconnection drive the rapid acceleration of the flux rope, coincident with the impulsive rise in the GOES soft X-ray flux. 

The close temporal correspondence between the Type III bursts and the ribbon intensity enhancements indicates that electron acceleration at the BCS produces bidirectional electron beams. The upward-moving beams generate Type III radio emission and in-situ electron fluxes, while the downward-propagating electrons precipitate and deposit energy in the chromosphere, producing localized brightenings in the circular ribbon and evaporation of excess coronal-temperature plasma that eventually may form coronal rain.

The inner HXR sources (S1 and S2), which coincide with the flare ribbons (Figure \ref{fig-xray}(c)), result from flare reconnection underneath the erupting flux rope.
Previous studies of EOVSA and STIX observations have suggested that the outer radio (microwave) and HXR sources originate from the footpoints of the erupting flux rope \citep{chen2020,stiefel2023}. In contrast, our results show that the sources in this event are clearly separated ($\approx$40 arcsec) from the flux-rope footpoints and are instead cospatial with the strong kernels at the circular ribbon produced during explosive breakout reconnection. Therefore we interpret the outer hard X-ray sources (Figure \ref{fig-xray}(c)) as signatures of breakout reconnection rather than emission from the footpoints of the erupting flux rope.  The northern source (S4) is comparatively fainter than the southern source (S3), indicating that most of the reconnection near the null occurred along leg L1. This process led to plasmoid formation, localized heating along L1, and the eventual disconnection of the leg. The enhanced brightness of S3 relative to S4 is consistent with the observed intensity asymmetry of the circular ribbon. The strongest kernels (S3, S4) of the circular ribbon and the associated HXR sources at the ends of the parallel ribbons likely mark preferred sites of particle precipitation associated with explosive reconnection near the null. These endpoints, often located near the legs of the erupting flux rope, correspond to regions where the magnetic connectivity converges and the squashing factor is high, resulting in concentrated energy deposition.

The photospheric magnetic connectivity maps of Parker Solar Probe (PSP) and near-Earth spacecraft reveal that both were magnetically connected to the erupting pseudostreamer (Figure \ref{app-fig2}). In contrast, STEREO-A was not directly connected to the pseudostreamer source region. Nevertheless, STEREO-A detected the associated SEP event, which is most plausibly explained by particle acceleration and subsequent injection over a broad longitudinal extent of the CME-driven shock front.

The velocity dispersion analysis suggests that the injection of electron beams occurs during the explosive phase of breakout reconnection (Figure \ref{app-fig4}). In this scenario, electrons accelerated at the null-point reconnection site propagate both downward and upward along newly reconnected field lines. The downward-propagating electron beams impact the lower atmosphere (i.e., chromosphere), producing the observed quasicircular flare ribbon and the external hard X-ray footpoint sources, while the upward-propagating electron beams escape along open or large-scale field lines, giving rise to the associated metric and decametric–hectometric Type~III radio bursts.

Previous observations have exhibited hard X-ray emission and Type III radio bursts associated with jets, highlighting the role of electron acceleration in evolving fan-spine magnetic configurations \citep{glesener2012,kumar2016a,chen2018,paipa-Leon2025}. In the present event, we observe a sustained storm of quasiperiodic Type III radio bursts accompanying interchange reconnection and the formation of pre-eruptive faint jets. Furthermore, NRH imaging localizes the Type III sources to the null-point topology, providing direct evidence that the interchange reconnection served as the source region for the escaping electron beams. Furthermore, the explosive BCS reconnection during the flux rope eruption observed in this event is in agreement with earlier observational and numerical studies suggesting that energetic electrons are initially confined within the erupting flux rope and are subsequently released onto open magnetic field lines during breakout reconnection, producing strong Type III bursts \citep{demoulin2012,bain2012,masson2019,li2025,kumar2025}. These similarities support the interpretation that BCS reconnection provides an efficient pathway for energetic particles to escape into the heliosphere.

The observed SEP profiles suggest contributions from two distinct but complementary acceleration and release processes. The initial enhancement (multiple bumps prior to the SEP peak) in the Wind/3DP electron and SOHO/ERNE proton intensity profiles likely corresponds to particles released promptly during the BCS reconnection, consistent with the onset of the Type III radio bursts and the establishment of magnetic connectivity between the erupting flux system and surrounding open magnetic field lines. While these enhancements may reflect multiple particle injection episodes, variations in particle transport within different magnetic flux tubes intersecting the spacecraft could also contribute to the observed intensity profile.  The subsequent gradual rise ($<$1 hr) and prolonged duration of the SEP event are more consistent with continuous acceleration by the expanding CME-driven shock (e.g., \citealt{vourlidas2013,kumar2025a}), which persists well beyond the limited interval during which the low-coronal EUV shock is visible. As the shock propagates through the outer corona and heliosphere, it can intercept progressively larger regions of the heliospheric magnetic field, sustaining particle acceleration and injection over extended timescales. This expanding shock geometry also provides a natural explanation for the broad longitudinal extent of the SEP event (e.g., \citealt{lario2016}), while interplanetary transport processes, including pitch-angle scattering and moderate cross-field diffusion, may further enhance the observed spread. The longitudinal spread of SEPs has also been reported in association with narrow SEP sources such as jets even in the absence of CME-driven shocks (e.g., \citealt{wiedenbeck2013, lario2024}), indicating that particles accelerated in interchange reconnection processes alone may considerably spread in the heliosphere and thus be observed by multiple spacecraft. 
 These results support a scenario in which the BCS primarily provides the early escape pathway for energetic electrons, whereas the CME-driven shock dominates the long-duration gradual SEP event and the widespread SEP distribution.

\subsection{MHD-wave mode conversion}
The emergence of a slower propagating disturbance above the initial null location shortly after the passage of the fast EUV wave suggests a physical connection between the shock–null interaction and the generation of the slow wavefront. The measured speed is comparable to the expected coronal sound speed
for a plasma temperature of 1.2~MK, which is consistent with its interpretation as a slow magnetoacoustic wave.
 When the fast-mode shock encounters the null-point topology, where the Alfv\'en speed decreases significantly, part of the compressive perturbation can be converted into a field-aligned slow-mode disturbance. The subsequent propagation of this slow wave along the separatrix and/or open magnetic field lines is consistent with theoretical predictions of MHD-wave mode conversion in the vicinity of magnetic nulls \citep{mcLaughlin2011}. The timing of the slow wave appearing only after the shock traverses the null region further supports the scenario that the slow-mode wave is a response to the shock-induced compression and restructuring of the fan-spine configuration. This result agrees well with earlier observational studies demonstrating mode conversion in null-point topologies \citep{kumar2024a,kumar2025}.

\section{CONCLUSION}\label{conclusion}
We present direct imaging of key signatures of interchange reconnection that generates recurrent jets at the open–closed boundary of a pseudostreamer. Interchange reconnection in pseudostreamer provides an efficient mechanism for transferring mass and energy from closed coronal structures into open magnetic field lines, thereby contributing to the solar wind. The recurrent jets produced during the reconnection episodes can supply heated plasma and Alfv\'enic perturbations to the open field, consistent with proposed sources of the slow solar wind. At the same time, the localized heating near the null point, followed by radiative cooling and the formation of coronal rain, indicates that reconnection-driven energy release is intermittent and spatially structured. These observations establish a direct observational link between interchange reconnection, jets, escaping electron beams (storm of quasiperiodic Type IIIs), plasma heating, and subsequent cooling/rain.

These results have important implications for solar wind structures. PSP provides unprecedented in situ measurements of magnetic and plasma fluctuations in the near-Sun environment, where the signatures of such coronal drivers are less processed by solar wind evolution. The observed coronal periodicities (5, 10 min) offer a potential remote-sensing diagnostic for identifying the coronal origin of microstreams and switchbacks detected by PSP \citep{kumar2023b}. Establishing this connection strengthens the interpretation that a fraction of small-scale solar wind structures may originate from quasiperiodic interchange reconnection at open-closed magnetic boundaries in pseudostreamers.

The pre-eruption jets and the opening of overlying flux is followed by the slow rise and subsequent eruption of the filament-carrying flux rope via breakout reconnection.
We identified hard X-ray and radio signatures associated with breakout reconnection in a pseudostreamer. The detection of four footpoint sources provides the first confirmation that the external sources result from the downward precipitation of electron beams released during explosive breakout reconnection. Simultaneous, intense Type III bursts reveal the upward escape of these electron beams, demonstrating a common origin through breakout reconnection. The observed energetic particle signatures are consistent with energization and injection of electrons from the breakout current sheet onto open field lines. In addition, the rapid acceleration of the erupting flux rope, associated with the removal of overlying ambient magnetic flux, leads to the formation of a shock in the low corona ahead of the flux rope \citep{kumar2025}. This shock can further accelerate energetic particles, including electrons and protons, contributing to the observed gradual SEP event. We also observe clear signatures of MHD-wave mode conversion as a fast-mode shock propagates through a neighboring fan-spine configuration, emphasizing the crucial role of magnetic topology in redistributing wave energy and facilitating localized plasma heating. We demonstrate the potential of multi-point observations from different missions in advancing our understanding of solar eruptions (jets and CMEs) and associated particle acceleration in pseudostreamers and similar flux systems. These results motivate further coordinated observations and MHD simulations to comprehend the wide range of dynamic activities associated with pseudostreamers.\\
\\
%%%%%%%%%%%%%%%%%%%%%%%%%%%%%%%%%%%%%%%%%%%%%%%%%%%%%%%%%%%%%%%%%%%%
%\begin{acknowledgments}
%We thank the referee for insightful comments that have improved this paper. 
\noindent
{\bf {Acknowledgements}}\\
SDO is a mission for NASA's Living With a Star (LWS) program. 
Parker Solar Probe was designed, built, and is now operated by the Johns Hopkins Applied Physics Laboratory as part of NASA’s Living with a Star (LWS) program (contract NNN06AA01C). Support from the LWS management and technical team has played a critical role in the success of the Parker Solar Probe mission. Thanks to the FIELDS team for providing data (PI: Stuart D. Bale, UC Berkeley). Thanks to the Integrated Science Investigation of the Sun (IS$_\sun$IS) Science Team (PI: David McComas, Princeton University). Thanks to the Solar Wind Electrons, Alphas, and Protons (SWEAP) team for providing data (PI: Justin Kasper, BWX Technologies). We thank the RSDB service at LESIA/USN (Observatoire de Paris) for making the NRH/ORFEES/NDA data available. We equally thank the radio monitoring service at LESIA (Observatoire de Paris) for providing value-added data used for this study. We acknowledge the USAF Radio Solar Telescope Network (RSTN), the e-Callisto network, and the Wind/WAVES instrument team for providing the solar radio observations used in this study. The Solar-MACH tool was originally developed at Kiel University, Germany and further discussed within the ESA Heliophysics Archives USer (HAUS) group.
Magnetic-field extrapolation was visualized with VAPOR (www.vapor.ucar.edu), a product of the Computational Information Systems Laboratory at the National Center for Atmospheric Research. Wavelet software was provided by C. Torrence and G. Compo, and is available at (\url{http://paos.colorado.edu/research/wavelets/}). K-Cor data are available at DOI: 10.5065/D69G5JV8 and the MLSO data portal (\url{https://mlso.hao.ucar.edu/mlso_data_calendar.php}).
 Courtesy of the Mauna Loa Solar Observatory, operated by the High Altitude Observatory, as part of the National Center for Atmospheric Research (NCAR). NCAR is supported by the National Science Foundation.
The CME catalog is generated and maintained at the CDAW Data Center by NASA and The Catholic University of America in cooperation with the Naval Research Laboratory. SOHO is a project of international cooperation between ESA and NASA.
We acknowledge the STEREO/IMPACT team and the Principal Investigator, Christina Lee (UCB/SSL), as well as the CDAWeb archive for the STEREO/IMPACT data.
Solar Orbiter is a mission of international cooperation between ESA and NASA. We acknowledge the use of data from the EUI and STIX onboard Solar Orbiter. EUI data are provided by the EUI consortium led by the Royal Observatory of Belgium. STIX is an international collaboration led by the University of Applied Sciences and Arts Northwestern Switzerland (FHNW).
 This research is supported by NSF SHINE Award (\#2229336), NASA's Heliophysics Guest Investigator (\#80NSSC25K7679) and Supporting Research (\#80NSSC24K0264) programs. PW was supported by a Leverhulme project grant (RPG-2023-288). The computations were sponsored by allocations on Discover at NASA's Center for Climate Simulation and on the DiRAC Data Analytic system at the University of Cambridge, operated by the University of Cambridge High Performance Computing Service on behalf of the STFC DiRAC HPC Facility (www.dirac.ac.uk) and funded by BIS National E-infrastructure capital grant (ST/K001590/1), STFC capital grants ST/H008861/1 and ST/H00887X/1, and STFC DiRAC Operations grant ST/K00333X/1. DiRAC is part of the National E-Infrastructure.\\
 \\

%\end{acknowledgments}
%%%%%%%%%%%%%%%%%%%%%%%%%%%%%%%%%%%%%%%%%%%%%%%%%%%%%%%%%%%%%%%%%%%%
%%%%%%%%%%%%%%%%%%%%%%%%%%%%%%%%%%%%%%%%%%%%%%%%%%%%%%%%%%%%%%%%%%%%

\bibliographystyle{aasjournal}
\bibliography{reference.bib}

\clearpage

\appendix
\counterwithin{figure}{section}
\section{Supplementary figures}

\begin{figure*}
\centering{
\includegraphics[width=14cm]{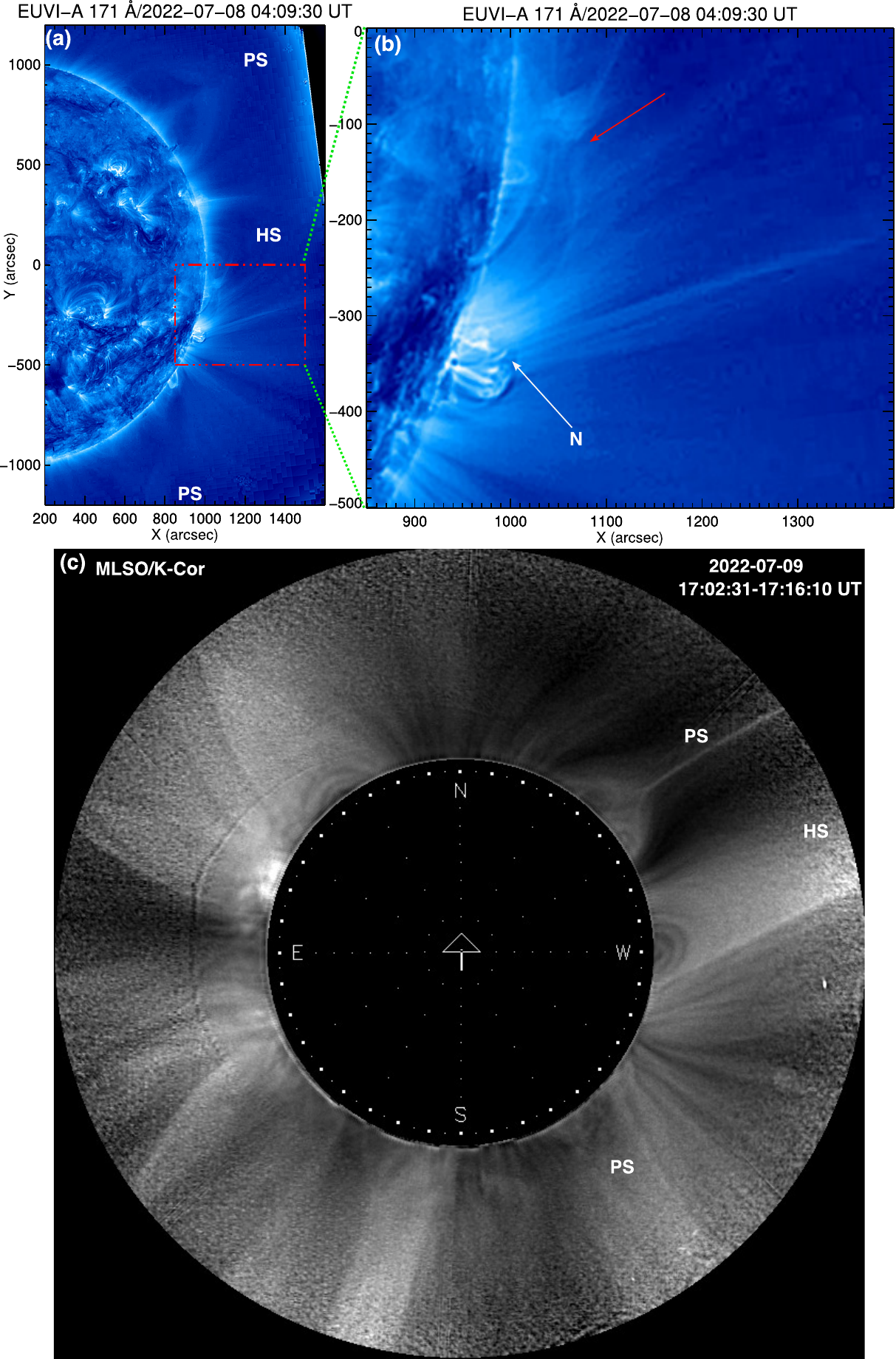}
}
\caption{Large-scale structures near the erupting pseudostreamer. 
(a, b) STEREO/EUVI-A 171~\AA~ images on July 8, 2022, showing large-scale helmet and pseudostreamers  (marked by HS and PS) in the corona. (b) A zoomed-in view of the null-point topology marked within the red rectangular box in (a). The red arrow indicates another closed fan–spine topology at the base of the HS.
(c) MLSO K-Cor coronagraph (enhanced intensity) image on July 9, 2022 (post-eruption phase). The dotted circle indicates the photosphere.} 
\label{app-fig1}
\end{figure*}

Figure \ref{app-fig1} shows an extended view of the eruption site (southernmost fan–spine topology) and nearby large-scale structures (HS and PS) in the STEREO/EUVI-A 171~\AA~ and Mauna Loa Solar Observatory’s (MLSO) K-coronagraph (K-Cor) images. K-Cor provides polarization brightness (pB) images of the white-light corona from 1.05 to 3.0 R$_\odot$, at 5.6$\arcsec$ pixel resolution.

\begin{figure*}
\centering{
\includegraphics[width=12cm]{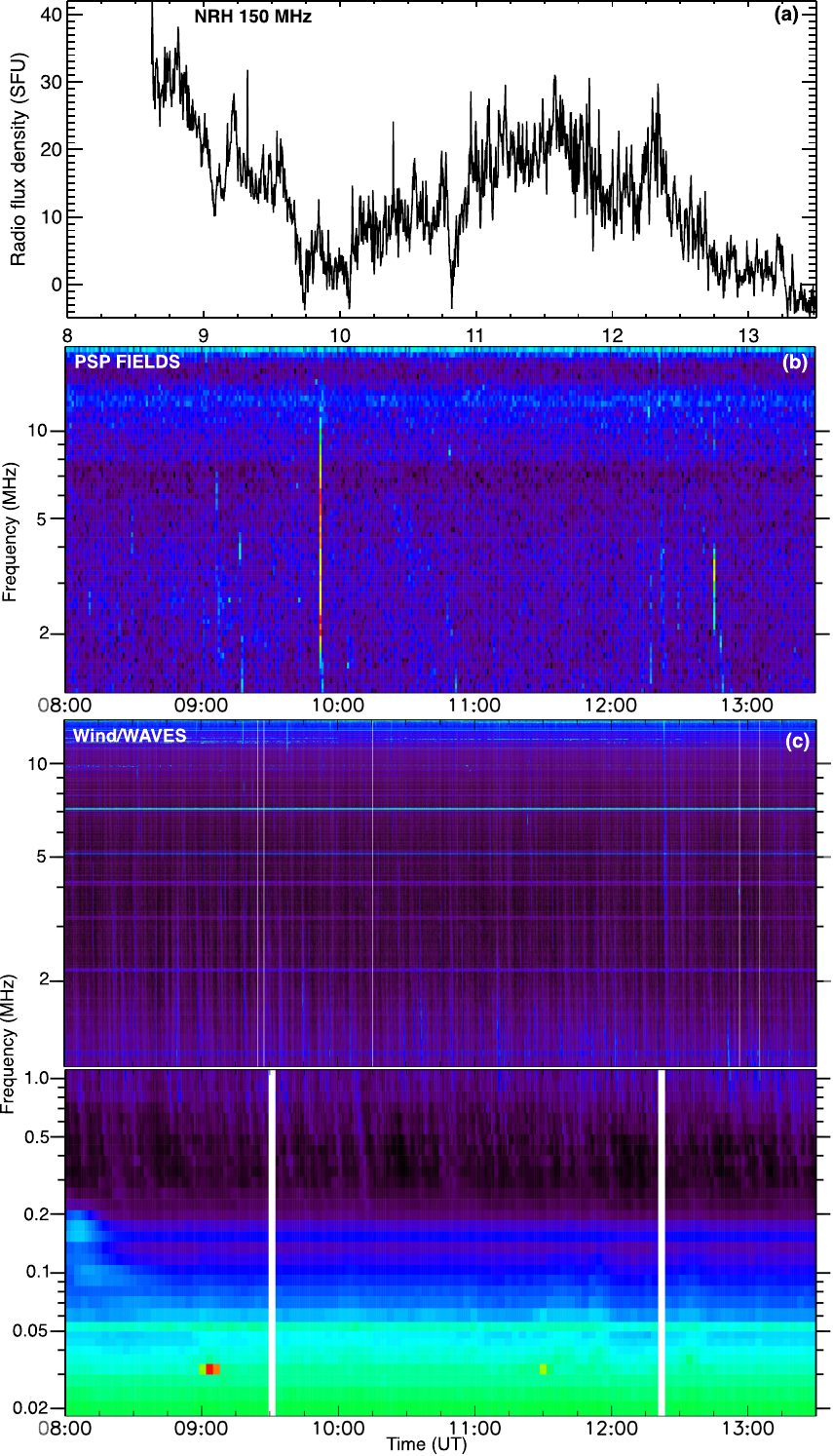}
}
\caption{Type III radio bursts observed during jet activity from the pseudostreamer on July 9, 2022.
(a) 150 MHz radio flux density extracted from the NRH radio source (Figure \ref{fig-radio}(a)) during the pre-eruption jets.
(b) Dynamic radio spectrum from PSP/FIELDS in the 1.3–19.2 MHz frequency range.
(c) Dynamic radio spectrum from Wind/WAVES in the 0.02–13.825 MHz frequency range.
} 
\label{app-fig3a}
\end{figure*}

\begin{figure*}
\centering{
\includegraphics[width=11cm]{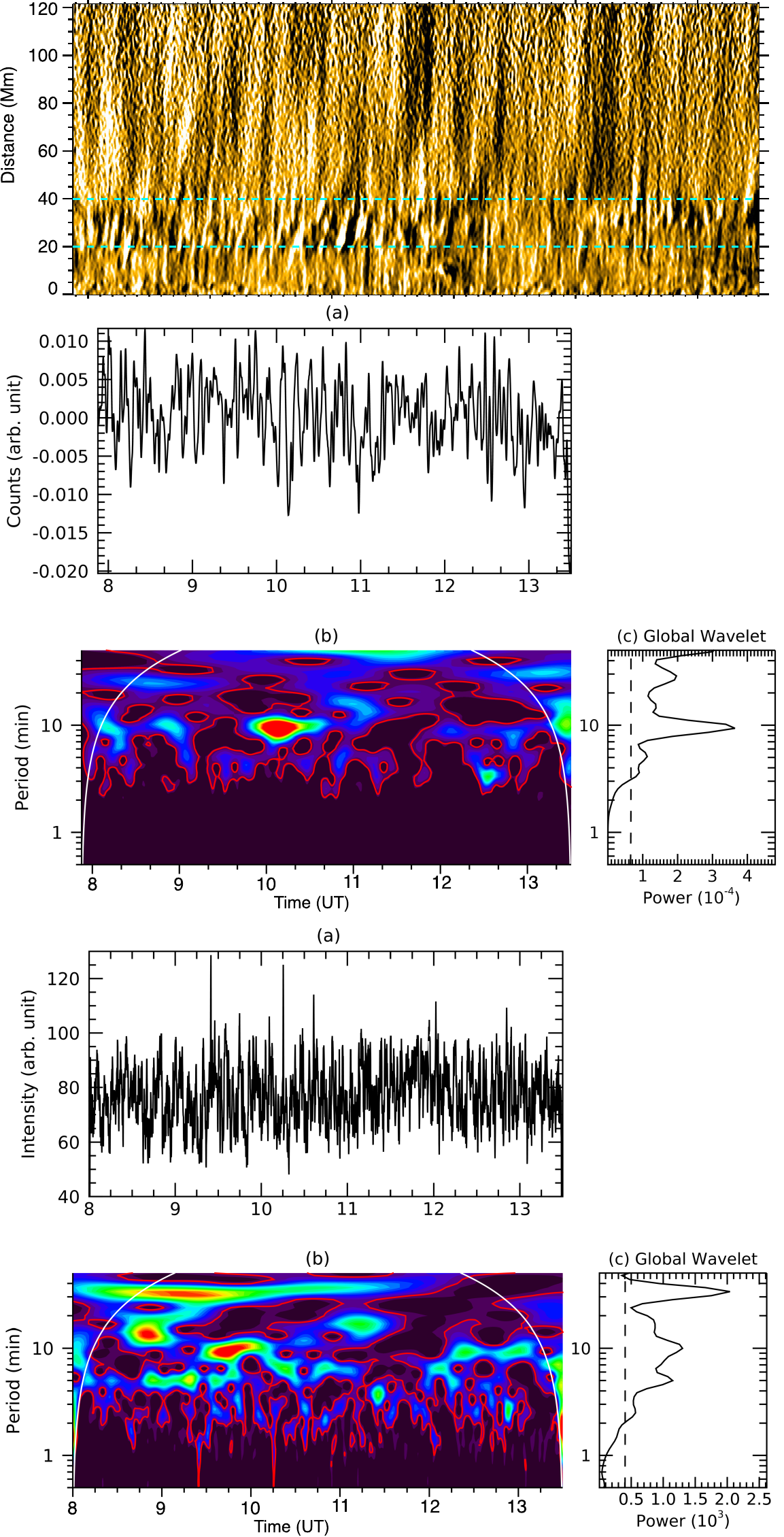}
}
\caption{{Quasiperiodic jets and associated Type III radio bursts from the pseudostreamer on July 9, 2022.}
\textit{Top:} Time--distance intensity (running-difference) plot along slice P4Q4 using AIA 171~\AA~ images.
\textit{Middle:} (a) Wavelet analysis of the intensities associated with recurrent jets, extracted from the time--distance plot between the two horizontal lines indicated in the top panel. 
(b) Wavelet power spectrum. Red contours indicate the 95\% significance level. 
(c) Global wavelet power spectrum. The dashed line represents the 95\% global confidence level.
\textit{Bottom:} Similar wavelet analysis for the Type~III radio bursts extracted from the Wind/WAVES dynamic spectrum at 1.5~MHz.
} 
\label{app-fig3b}
\end{figure*}

\begin{figure*}
\centering{
\includegraphics[width=12cm]{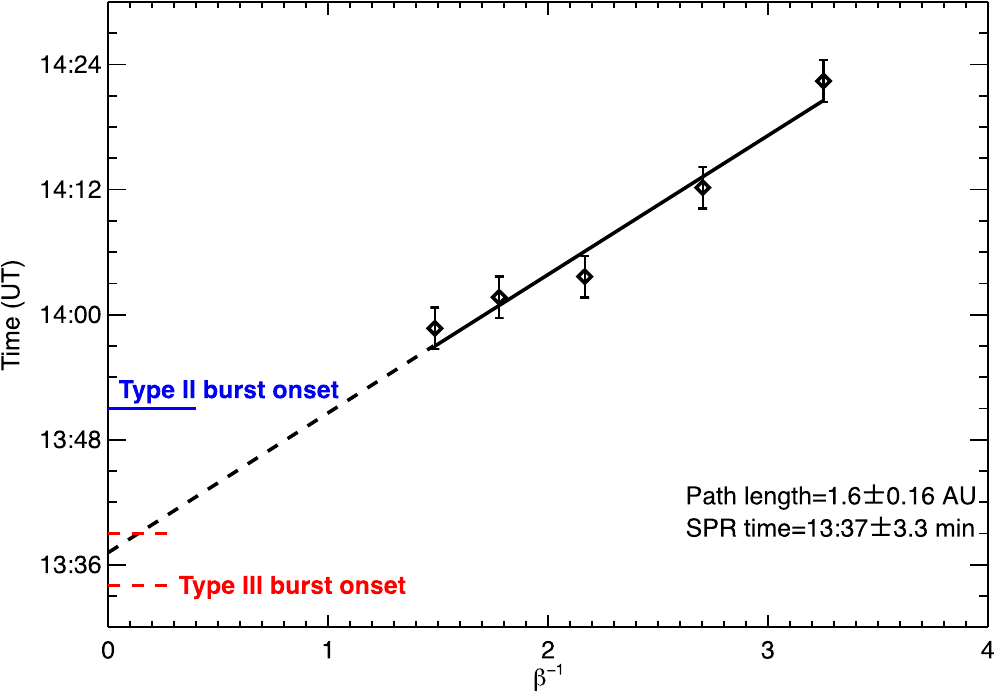}
}
\caption{Velocity dispersion analysis for the onset of SEP (electrons) for the on July 9, 2022. Inverse beta vs onset time plot for electrons (26–180 keV) detected by Wind/3DP at 1 AU. The red dashed lines indicate the onset and end times of metric Type III bursts, while the blue line marks the onset time of the metric Type II radio burst. SPR denotes solar particle release.
} 
\label{app-fig4}
\end{figure*}

The low coronal (150 MHz) and interplanetary Type III bursts observed during the quasiperiodic jets from the pseudostreamer are shown in Figure \ref{app-fig3a}. Wavelet analyses of the quasiperiodic jets and associated Type IIIs were performed to estimate their periodicity (Figure \ref{app-fig3b}). The wavelet analysis is applied to the original signal. No additional detrending was applied. The wavelet transform was computed using the Morlet mother wavelet \citep{torrence1998}. Statistical significance was evaluated at the 95$\%$ confidence level using the default white-noise background spectrum implemented in the wavelet code. The cone of influence (COI) is shown in the wavelet spectrum, and only periodicities inside the COI are considered reliable.
Figure~\ref{app-fig4} shows the results of the velocity dispersion analysis for electrons (Wind/3DP).

%%%%%%%%%%%%%%%%%%%%%%%%%%%%%%%%%%%%%%%%%%%%%%%%%%%%%%%%%%%%%%
%%%%%%%%%%%%%%%%%%%%%%%%%%%%%%%%%%%%%%%%%%%%%%%%%%%%%%%%%%%%%%%

%%%%%%%%%%%%%%%%%%%%%%%%%%%%%%%%%%%%%%%%%%%%%%%%%%%%%%%%%%%%%%
%%%%%%%%%%%%%%%%%%%%%%%%%%%%%%%%%%%%%%%%%%%%%%%%%%%%%%%%%%%%%%%
%%%%%%%%%%%%%%%%%%%%%%%%%%%%%%%%%%%%%%%%%%%%%%%%%%%%%%%%%%%%%%
%%%%%%%%%%%%%%%%%%%%%%%%%%%%%%%%%%%%%%%%%%%%%%%%%%%%%%%%%%%%%%%
%%%%%%%%%%%%%%%%%%%%%%%%%%%%%%%%%%%%%%%%%%%%%%%%%%%%%%%%%%%%%%
%%%%%%%%%%%%%%%%%%%%%%%%%%%%%%%%%%%%%%%%%%%%%%%%%%%%%%%%%%%%%%%

\section{Supplementary Materials}
This section contains supplementary movies to support the results. All supplementary movies are available in the Zenodo repository at doi:\href{https://doi.org/10.5281/zenodo.22044302}{10.5281/zenodo.22044301}. \\
{\bf Movie S1}: An animation of the AIA 304~{\AA}, 171~{\AA} and 193~{\AA} images during the pre-eruption phase (Figure \ref{fig-pre}). The animation runs from 08:01:23 UT to 13:30:11 UT. Its real-time duration is 16.6 s. \\
{\bf Movie S2}: An animation of the AIA 131~{\AA} images during the flux-rope eruption (Figure \ref{fig-aia1}). The animation runs from 13:17:47 UT to 14:01:47 UT. Its real-time duration is 5 s. \\
{\bf Movie S3}: An animation of the AIA 1600~{\AA} intensity and running-difference images during the eruption (Figure \ref{fig-aia1}). The animation runs from 13:20:35 UT to 13:59:23 UT. Its real-time duration is 3.9 s. \\
{\bf Movie S4}: An animation of the AIA 171~{\AA} intensity and running-difference images overlaid by NRH radio contours (150 MHz) during the pre-eruption and eruption phases (Figure \ref{fig-radio}). The animation runs from 13:32:59 UT to 14:40:11 UT. Its real-time duration is 3.7 s. \\
{\bf Movie S5}: An animation of the AIA 193~{\AA} intensity/running-difference images and time-distance intensity plots along the slices R1S1 and R2S2 (Figure \ref{fig-mode}). The animation runs from 13:17:47 UT to 14:02:35 UT. Its real-time duration is 4.5 s. \\
{\bf Movie S6}: An animation of the 3D MHD simulation of a pseudostreamer eruption (Figure \ref{fig-sim}). Its real-time duration is 2.1 s. \\

%%%%%%%%%%%%%%%%%%%%%%%%%%%%%%%%%%%%%%%%%%%%%%%%%%%%%%%%%%%%%%%

\end{document}